\documentclass[a4paper,11pt]{article}
\pdfoutput=1
\usepackage{jheppub}

\usepackage{amsmath,bm,amssymb,amsthm,mathrsfs,mathtools}
\usepackage{subcaption,color}
\usepackage{ragged2e}
\usepackage{caption}
\usepackage{array}
\justifying\let\raggedright\justifying

\allowdisplaybreaks[4]
\usepackage{tikz}
\usepackage{multirow}
\usepackage{comment}
\usetikzlibrary{backgrounds}
\usetikzlibrary{shapes,matrix,trees}
\usetikzlibrary{arrows.meta}
\usepackage{centernot}
\usetikzlibrary{positioning}			% For "above of=" commands
\usetikzlibrary{calc,through}			% For coordinates
\usetikzlibrary{decorations.pathreplacing}     % For curly braces
\usepackage{pgffor}                                        % For repeating patterns
\usetikzlibrary{decorations.pathmorphing}	% For Feynman Diagrams
\usetikzlibrary{decorations.markings}
\tikzset{
	vector/.style={decorate, decoration={snake}, draw},
	provector/.style={decorate, decoration={snake,amplitude=2.5pt}, draw},
	antivector/.style={decorate, decoration={snake,amplitude=-2.5pt}, draw},
	fermion/.style={draw=black, postaction={decorate},
		decoration={markings,mark=at position .55 with {\arrow[draw=black]{>}}}},
	fermionbar/.style={draw=black, postaction={decorate},
		decoration={markings,mark=at position .55 with {\arrow[draw=black]{<}}}},
	fermionnoarrow/.style={draw=black},
	gluon/.style={decorate, draw=black,
		decoration={coil,amplitude=4pt, segment length=5pt}},
	scalar/.style={dashed,draw=black, postaction={decorate},
		decoration={markings,mark=at position .55 with {\arrow[draw=black]{>}}}},
	scalarbar/.style={dashed,draw=black, postaction={decorate},
		decoration={markings,mark=at position .55 with {\arrow[draw=black]{<}}}},
	scalarnoarrow/.style={dashed,draw=black},
	electron/.style={draw=black, postaction={decorate},
		decoration={markings,mark=at position .55 with {\arrow[draw=black]{>}}}},
	bigvector/.style={decorate, decoration={snake,amplitude=4pt}, draw},
	photon/.style={decorate, draw=black,decoration={snake,amplitude=4pt, segment length=5pt} }
}

\definecolor{ccblue}{rgb}{0.0,0.4,0.8}
\usepackage[]{hyperref}
\hypersetup{  colorlinks=true,
	linkcolor=ccblue,
	urlcolor=ccblue,
	citecolor=ccblue}
\usepackage{amsmath}
\usepackage{amsfonts}
\usepackage{amssymb}
\usepackage{bbm}
\usepackage{booktabs}
\usepackage{slashed}
\usepackage[]{hyperref}
\usepackage{autobreak}
\newcommand{\tr}{\mathrm{Tr}}
\newcommand{\trans}{\mathrm{T}}
\preprint{PSI-PR-23-25,~ZU-TH-35/23}

\title{\boldmath Renormalization Group Evolution with Scalar Leptoquarks
}
\author[a,b]{Sumit Banik}
\author[a,b]{Andreas Crivellin}
\affiliation[a]{Physik-Institut, Universit\"at Z\"urich, Winterthurerstrasse 190, CH--8057 Z\"urich, Switzerland}
\affiliation[b]{Paul Scherrer Institut, CH--5232 Villigen PSI, Switzerland}

\emailAdd{sumit.banik@psi.ch}
\emailAdd{andreas.crivellin@psi.ch}

\abstract{Leptoquarks are theoretically well-motivated and have received increasing attention in recent years as they can explain several hints for physics beyond the Standard Model. In this article, we calculate the renormalisation group evolution of models with scalar leptoquarks. We compute the anomalous dimensions for all couplings (gauge, Yukawa, Higgs and leptoquarks interactions) of the most general Lagrangian at the two-loop level and the corresponding threshold corrections at one-loop. The most relevant analytic results are presented in the Appendix, while the notebook containing the full expressions can be downloaded at \href{https://github.com/SumitBanikGit/SLQ-RG}{https://github.com/SumitBanikGit/SLQ-RG}. In our phenomenological analysis, we consider some exemplary cases with focus on gauge and Yukawa coupling unification.}
\newpage

\begin{document} 
\maketitle

\section{Introduction}\label{sec1}
The Standard Model (SM) of particle physics describes the known fundamental constituents of matter as well as their interactions. While the Higgs particle~\cite{Higgs:1964ia,Englert:1964et,Higgs:1964pj,Guralnik:1964eu}, last missing puzzle piece of the SM, was discovered in 2012 at the Large Hadron Collider (LHC) at CERN~\cite{Aad:2012tfa,Chatrchyan:2012ufa,CDF:2012laj}, it is clear that the SM cannot be the ultimate theory of Nature: It does not account for the astrophysical observation of Dark Matter nor for the non-vanishing neutrino masses required by neutrino oscillations. 

A plethora of possible SM extensions have been proposed in the last decade. In this context, leptoquarks (LQs), i.e.~hypothetical new particles which directly couple a quark to a lepton, are very interesting. They were first proposed within Grand Unified Theories (GUTs)~\cite{Pati:1974yy,Fritzsch:1974nn,Georgi:1974sy,Georgi:1974yf} and composite models~\cite{Pati:1975md,Wudka:1985ef,Eboli:1987vb,Gripaios:2009dq}, and squarks can act as LQs within the R-parity violating MSSM (see e.g. Ref.~\cite{Barbier:2004ez} for a review). They were first classified in Ref.~\cite{Buchmuller:1986zs} into ten possible representations under the SM gauge group, of which five are scalars, and five are vector particles. 

In recent years, there has been renewed interest in LQs as they give interesting effects in low energy observables in general~\cite{Shanker:1981mj,Shanker:1982nd,Leurer:1993em,Leurer:1993qx,Davidson:1993qk,Crivellin:2021bkd,Crivellin:2020mjs}, and could provide solutions to anomalies in semi-leptonic $B$ decays~\cite{Gripaios:2014tna,Alonso:2015sja, Calibbi:2015kma, Fajfer:2015ycq, Bhattacharya:2016mcc, Buttazzo:2017ixm, Barbieri:2015yvd, Barbieri:2016las, Calibbi:2017qbu, Bordone:2017bld, Bordone:2018nbg, Kumar:2018kmr, Biswas:2018snp, DaRold:2018moy, Crivellin:2018yvo, Blanke:2018sro, Heeck:2018ntp,deMedeirosVarzielas:2019lgb, Cornella:2019hct, Bordone:2019uzc,Sahoo:2015wya, Chen:2016dip, Dey:2017ede, Becirevic:2017jtw, Chauhan:2017ndd, Becirevic:2018afm, Popov:2019tyc,Fajfer:2012jt, Deshpande:2012rr, Freytsis:2015qca, Bauer:2015knc, Li:2016vvp, Zhu:2016xdg, Popov:2016fzr, Deshpand:2016cpw, Becirevic:2016oho, Cai:2017wry, Altmannshofer:2017poe, Kamali:2018fhr, Mandal:2018kau, Azatov:2018knx, Wei:2018vmk, Angelescu:2018tyl, Kim:2018oih, Aydemir:2019ynb, Yan:2019hpm,Crivellin:2017zlb, Marzocca:2018wcf, Bigaran:2019bqv,Crivellin:2019dwb,Saad:2020ihm,Dev:2020qet,Saad:2020ucl,Altmannshofer:2020axr,Fuentes-Martin:2020bnh,Gherardi:2020qhc,DaRold:2020bib,Aebischer:2022oqe,Hiller:2016kry, Crivellin:2017dsk, Crivellin:2019szf, Bernigaud:2019bfy,Aebischer:2018acj,Fuentes-Martin:2019ign,Varzielas:2015iva,Sahoo:2016pet,Fajfer:2018bfj,Aydemir:2018cbb,Carvunis:2021dss,Zhang:2021dgl,Marzocca:2021miv,Barbieri:2022ikw,Garcia-Duque:2022tti,Crivellin:2022mff,FernandezNavarro:2022gst,Ismael:2023ffx,Iguro:2023prq,Lizana:2023kei},
the anomalous magnetic moment of the muon ($a_\mu$)~\cite{Bauer:2015knc,Djouadi:1989md, Chakraverty:2001yg,Cheung:2001ip,Popov:2016fzr,Chen:2016dip,Biggio:2016wyy,Davidson:1993qk,Couture:1995he,Mahanta:2001yc,Queiroz:2014pra,ColuccioLeskow:2016dox,Chen:2017hir,Das:2016vkr,Crivellin:2017zlb,Cai:2017wry,Crivellin:2018qmi,Kowalska:2018ulj,Dorsner:2019itg,Crivellin:2019dwb,DelleRose:2020qak,Saad:2020ihm,Bigaran:2020jil,Dorsner:2020aaz,Fuentes-Martin:2020bnh,Gherardi:2020qhc,Babu:2020hun,Aebischer:2021uvt,Crivellin:2020tsz,Bigaran:2022kkv,Stockinger:2022ata,Parashar:2022wrd,Chen:2022hle} in particular. The effects of LQs at colliders~\cite{Djouadi:1989md,Kramer:1997hh,Kramer:2004df,Faroughy:2016osc,Greljo:2017vvb,Blumlein:1996qp, Dorsner:2017ufx, Cerri:2018ypt,Bandyopadhyay:2018syt,Hiller:2018wbv,Faber:2018afz,Schmaltz:2018nls,Chandak:2019iwj,Allanach:2019zfr,Buonocore:2020erb,Haisch:2020xjd,Borschensky:2020hot,Crivellin:2021egp,Nomura:2021oeu,Ban:2021tos,FileviezPerez:2021lkq,Wang:2021uqz,Cheung:2022zsb,Ghosh:2022vpb,Borschensky:2022xsa,Dorsner:2022ibm}, in oblique electroweak parameters as well as Higgs couplings to gauge bosons~\cite{Keith:1997fv,Dorsner:2016wpm,Bhaskar:2020kdr,Zhang:2019jwp,Gherardi:2020det,Crivellin:2020ukd,Albergaria:2023pgl,Marzo:2022nrw}, electric dipole moments~\cite{Dekens:2018bci,Crivellin:2019qnh} and proton decay~\cite{Dorsner:2012nq,Dorsner:2022twk} were studied. LQs were considered as a portal to dark matter~\cite{Manzari:2022iyn,Queiroz:2014pra,Mandal:2018czf,Belanger:2022kvj,Belfatto:2021ats,Belanger:2021smw,Singirala:2021gok,Sahoo:2021vug,Mohamadnejad:2019wqb} and as the origin of neutrino masses~\cite{Mahanta:1999xd,Deppisch:2016qqd,Popov:2016fzr,Dorsner:2017wwn,Cheung:2017efc,Cai:2017wry,Saad:2020ucl,Babu:2020hun,Chen:2022hle}. Furthermore, they have been searched for at the LHC~\cite{ATLAS:2023prb,ATLAS:2023vxj,ATLAS:2023kek,ATLAS:2023uox,ATLAS:2022wcu,ATLAS:2021oiz,ATLAS:2020xov,ATLAS:2020dsk,ATLAS:2019qpq,ATLAS:2019ebv,CMS:2022goy,CMS:2020wzx,CMS:2018yiq,CMS:2018ncu,CMS:2018iye,CMS:2018oaj,CMS:2018lab,CMS:2018txo,CMS:2018qqq,CMS:2018svy} resulting in bounds of the order of $1-2\,$TeV (in the absence or suppression of couplings to first generation quarks~\cite{Crivellin:2021egp,Crivellin:2021bkd}).

As LQs are possible (light) remnants of a GUT~\cite{Preda:2022izo}, it is interesting to assess their impact on gauge and Yukawa coupling unification. In particular, scalar LQs (SLQs) could be light states of a GUT symmetry-breaking sector~\cite{Becirevic:2018afm} and it might even be possible that they are the only new particles (in addition to the SM) up to the GUT scale. In this case, they must alter the (renormalization group evaluation) RGE sufficiently to lead to the required coupling unification~\cite{Murayama:1991ah}. 

In this article, we will analyze the RGE in models with SLQs \footnote{As for vector LQs a Higgs mechanism is necessary to make the model renormalizable, one can not study the RGEs in a simplified model approach.}. For this, we will calculate the two-loop anomalous dimensions as well as the one-loop threshold corrections (at the LQ scale) and apply these results to the study of gauge and Yukawa coupling unification. For this, the article is structured as follows: In Sec.~\ref{sec2}, we present our setup and conventions. In Sec.~\ref{sec3}, we derive the $\beta$-function, give their analytic expression for some simple cases and examine gauge and Yukawa coupling unification for some specific examples. In Sec.~\ref{sec4}, we summarize and discuss the results of this paper. In Appendix \hyperlink{AppLQLag}{A}, we give the full SLQ lagrangian, including all five possible SLQs. In Appendix \hyperlink{AppRGEGauge}{B} and \hyperlink{AppRGEYuk}{C}, we give the anomalous dimensions for the gauge (two-loop) and Yukawa couplings (one-loop), respectively. In Appendix \hyperlink{AppThreshold}{D}, we collect the one-loop threshold corrections of SM parameters on matching SLQ models with the SM. A notebook with our full results, including Higgs and LQ (self-) interactions etc., is available at \cite{GitHubSLQ}.

\section{Setup and Conventions}\label{sec2}
\begin{table}[h]
\begin{center}
\begin{tabular}{ | m{3.7 cm} | m{1.6cm}| m{1.6cm} | m{1.6cm}| m{1.6cm}| m{1.6cm}| m{1.6cm}| } 
  \hline
 SM fields & $Q_j$ & $L_j$ & $u_j$ & $d_j$ & $\ell_j$ & $H$ \\
  \hline
  $ U(1)_Y, SU(2)_{L}, SU(3)_c $ & $+\frac{1}{6}\,,\mathbf{2}\,,\mathbf{3}$ & $ - \frac{1}{2} \, ,\mathbf{2} \, ,\mathbf{1} $ & $ +\frac{2}{3} \, ,\mathbf{1} \, ,\mathbf{3} $ & $ - \frac{1}{3} \, ,\mathbf{1} \, ,\mathbf{3}$ & $ -1 \, ,\mathbf{1} \, ,\mathbf{1}$ & $ +\frac{1}{2} \, ,\mathbf{2} \, ,\mathbf{1}$ \\
  \hline

 Leptoquark & $\Phi_1$ & $\Phi_{\tilde{1}}$ & $\Phi_{2}$ & $\Phi_{\tilde{2}}$ & $\Phi_{3}$ & \\
  \hline
  $ U(1)_Y, SU(2)_{L}, SU(3)_c $ & $-\frac{1}{3}\,, \mathbf{1}\, ,\mathbf{3}$ & $-\frac{4}{3}\, ,\mathbf{1}\, ,\mathbf{3}$ & $+\frac{7}{6} \, ,\mathbf{2} \, ,\mathbf{3}$ & $+\frac{1}{6} \, ,\mathbf{2} \, ,\mathbf{3}$ & $- \frac{1}{3} \, ,\mathbf{3} \, ,\mathbf{3}$ &  \\
  \hline
\end{tabular}
\caption{Representations of the SM and LQ fields under the $SU(3)_c \times SU(2)_{L} \times U(1)_Y$ gauge group.}
\label{SMfields}
\end{center}
\end{table}

The SM fields transform under the $SU(3)_c \times SU(2)_{L} \times U(1)_Y$ gauge group as given in Table~\ref{SMfields}, where $j= 1,2,3 $ is a flavour index. The left-handed fermions $Q_j$ and $L_j$ are doublets under $SU(2)_L$ and decompose into their components as
\begin{equation}
Q_f = \begin{pmatrix}
  u_{j,L}\\ 
  d_{j,L}
\end{pmatrix} , \hspace{1cm}
L_f = \begin{pmatrix}
  \nu_{j,L}\\ 
  \ell_{j,L}
\end{pmatrix} \,,
\end{equation}
while the right-handed fields $u_j$, $d_j$ and $\ell_j$ are $SU(2)_L$ singlets which we write as
\begin{equation}
u_j = \begin{pmatrix}
  u_{j,R} 
\end{pmatrix} , \hspace{1cm}
d_j = \begin{pmatrix}
  \nu_{j,R}\\ 
\end{pmatrix} , \hspace{1cm}
\ell_j = \begin{pmatrix}
  \ell_{j,R}
\end{pmatrix}\,.
\end{equation}
The electric charge $q$ is given by the Gell-Mann-Nishijima formula, $q= Y + I_3$ where, $I_3$ is the third-component of the weak isospin and $Y$ is the hypercharge related to the gauge factor $U(1)_Y$. 

The SM Lagrangian is then written as
\begin{align}\label{LSM}
\mathcal{L}_{\mathrm{SM}}= & -\frac{1}{4} B^{\mu \nu} B_{\mu \nu} -\frac{1}{4} W^{I, \mu \nu} W_{I, \mu \nu}-\frac{1}{4} G^{\alpha, \mu \nu} G_{\alpha, \mu \nu} 
\nonumber \\ & 
 + i \left(\bar{Q}_{f}\slashed{D} Q_{f}\right)
 + i \left(\bar{L}_{f}\slashed{D} L_{f}\right)
 + i \, \bar{u}_{f}\slashed{D} u_{f}
 + i \, \bar{d}_{f}\slashed{D} d_{f}
 + i \, \bar{\ell}_{f} \slashed{D} \ell_{f}
 \\ & 
 +\left(\left(D^{\mu} H\right)^{\dagger} D_{\mu} H\right)+\mu_{H}^{2}\left(H^{\dagger} H\right)-\lambda_{H}\left(H^{\dagger} H\right)^{2}
\nonumber \\ & \nonumber
 -\left(Y_{f_i f_j}^{d}\left(\bar{Q}_{f_i} H\right) d_{f_j}+Y_{f_i f_j}^{\ell}\left(\bar{L}_{f_i} H\right) \ell_{f_j}+Y_{f_i f_j}^{u}\left(\bar{Q}_{f_i} i \sigma_2 H^{\dagger}\right) u_{f_j}+{\rm h.c.}\right)
 \nonumber \\ & \nonumber
 + \mathcal{L}_{\mathrm{gauge-fixing}}\,,
\end{align}
where $Y^d$, $Y^\ell$ and $Y^u$ are the Yukawa couplings and $B^{\mu \nu}$ and the field strength tensors defined as,
\begin{align} 
& \nonumber B_{\mu \nu} = \partial_\mu B_\nu - \partial_\nu B_\mu ,
\\ & 
W^{I}_{ \mu \nu} = \partial_\mu W^{I}_\nu - \partial_\nu W^{I}_\mu + g_2 f^{I J K} W^{J}_\mu W^{K}_\nu, \hspace{1cm} I \in \{ 1,2,3 \} \,,
\\ & \nonumber 
G^{\alpha}_{ \mu \nu} = \partial_\mu G^{\alpha}_\nu - \partial_\nu G^{\alpha}_\mu + g_3 f^{\alpha \beta \gamma} G^{\beta}_\mu G^{\gamma}_\nu\,, \hspace{1.5cm} \alpha \in \{ 1,..,8 \}\,,
\end{align} 
corresponding to the gauge groups $U(1)_Y$, $SU(2)_L$ and $SU(3)_c$, respectively. The structure constants of $SU(2)_L$ and $SU(3)_c$ are denoted by $f^{I J K}$ and $f^{\alpha \beta \gamma}$. For better readability, $SU(2)_L$ indices inside parenthesis $( \cdots )$ are contracted such that they form gauge singlets. The covariant derivative for any field $\phi$ is defined as
\begin{equation}
D_\mu \phi = \partial_\mu \phi - i \tilde{g}_1 Y B_\mu \phi -i g_2 \tau_I W^I_\mu \phi -i g_3 T_\alpha G^\alpha_\mu \phi\,,
\end{equation}
where $Y$ is the $U(1)_Y$ hypercharge of $\phi$. $\tilde{g}_1$, $g_2$ and $g_3$ are the gauge couplings associated with $U(1)_Y$, $SU(2)_L$ and $SU(3)_c$ gauge groups, respectively. $\tau_I$ and $T_\alpha$ are the generators of $SU(2)_L$ and $SU(3)_c$ depending on the representation of $\phi$. In the fundamental representation of $SU(2)_L$ and $SU(3)_c$ we have  $\tau_I=\frac{\sigma_I}{2}$  and $T_\alpha=\frac{\lambda_\alpha}{2}$, where $\sigma_I$ are the  Pauli matrices
\begin{align}
\sigma_1 = \begin{pmatrix}
0 & 1 \\
1 & 0 \\
\end{pmatrix}, \hspace{0.3cm}
\sigma_2 = \begin{pmatrix}
0 & -i \\
i & 0 \\
\end{pmatrix}, \hspace{0.3cm}
\sigma_3 = \begin{pmatrix}
1 & 0 \\
0 & -1 \\
\end{pmatrix} \,, 
\end{align}
 and $\lambda_\alpha$ are the Gell-Mann matrices.

We now add to the SM the five scalar SLQs which transform under the SM gauge groups as given in Table~\ref{SMfields}. For later convenience, we decompose the SLQ Lagrangian as
\begin{align} \label{LLQ}
\mathcal{L}_{\mathrm{LQ}}=&\hspace{0.3cm}\mathcal{L}_{1}+\mathcal{L}_{{\tilde{1}}}+\mathcal{L}_{2}+\mathcal{L}_{{\tilde{2}}}+\mathcal{L}_{3}+\mathcal{L}_{1\,{\tilde{1}}}+\mathcal{L}_{1\,{2}}+\mathcal{L}_{1\,{\tilde{2}}}+\mathcal{L}_{1\,{3}} 
\nonumber \\ & 
\hspace{-0.15cm} + \mathcal{L}_{{\tilde{1}}\,{2}}+\mathcal{L}_{{\tilde{1}}\,{\tilde{2}}}+\mathcal{L}_{{\tilde{1}}\,{3}}+\mathcal{L}_{{2}\,{\tilde{2}}}+\mathcal{L}_{{2}\,{3}}+\mathcal{L}_{{\tilde{2}}\,{3}}
\nonumber \\ &
\hspace{-0.15cm} +\mathcal{L}_{{\tilde{1}}\,{2}\,{\tilde{2}}} +\mathcal{L}_{{1}\,{\tilde{1}}\,{2}}+\mathcal{L}_{{1}\,{2}\,{3}}+\mathcal{L}_{{1}\,{\tilde{2}}\,{3}}+\mathcal{L}_{{\tilde{1}}\,{2}\,{3}}
\nonumber \\ &
\hspace{-0.15cm} +\mathcal{L}_{{\tilde{1}}\,{2}\,{\tilde{2}}\,{3}}+\mathcal{L}_{{1}\,{\tilde{1}}\,{2}\,{\tilde{2}}}\,,
\end{align}
where, the terms $\mathcal{L}_{i j ... }$ contain terms involving only the LQs $\{\Phi_i, \Phi_j, ... \}$. The explicit expressions, as already presented in Ref.~\cite{Crivellin:2021ejk}, for each Lagrangian term are given in Appendix~\hyperlink{AppLQLag}{A}. Note that for our purpose, it is imperative to have the complete Lagrangian \footnote{{We disregard $B$-number violating couplings as they must be small due to constraints from proton decay and therefore will have a small effect on RGEs.}}, e.g.~terms involving LQ-Higgs interactions, such that the full set of counterterms needed to cancel all generated divergences is available.

\section{Renormalization Group Evolution and Phenomenological Analysis}\label{sec3}

We calculate the complete two-loop $\beta$-functions and the one-loop threshold corrections of all couplings of our LQ Lagrangian in the $\overline{\text{\small MS}}$-scheme. For the anomalous dimensions, we used the Python package \texttt{PyR@TE}~\cite{Sartore:2020gou}.\footnote{We cross-checked our results using the \textit{Mathematica} package \texttt{RGBeta}~\cite{Thomsen:2021ncy}.} The \texttt{PyR@TE} model files and full analytic expressions for the $\beta$-functions, as well as the threshold corrections, are available online publicly on Github~\cite{GitHubSLQ}. The one-loop threshold corrections are presented in Appendix~\hyperlink{AppThreshold}{D}, which were derived using \texttt{Matchete}~\cite{Fuentes-Martin:2022jrf} and cross-checked for some selected cases with \texttt{matchmakereft}~\cite{Carmona:2021xtq}. We also checked that including the threshold corrections reduce, as expected, the matching scale dependence. The QCD contribution to some LQ Yukawa couplings was cross-checked with Ref.~\cite{Plehn:1997az}.

We use the following convention throughout this paper for the $\beta$-function for any parameter $A$ 
\begin{equation}
\beta\left(A\right) \equiv \mu \frac{d A}{d \mu}\equiv\frac{1}{\left(4 \pi\right)^{2}}\beta^{(1)}(A)+\frac{1}{\left(4 \pi\right)^{4}}\beta^{(2)}(A)+...\,.
\end{equation}
where $\mu$ is the renormalization scale and $\beta^{(n)}(A)$ denotes the contribution at $n$-loop order and the dots are the higher-order terms. The $\beta$-functions form a system of coupled differential equations which we solve numerically, as explained below. 

At the one-loop level, we have the following coefficients for the gauge couplings
\begin{align}
		\beta^{(1)}({g}_1) &=\frac{41}{10} {g}^3_1+\left(  \frac{n_1}{15} + \frac{16 n_{\tilde{1}}}{15} + \frac{49 n_2}{30} + \frac{n_{\tilde{2}}}{30} + \frac{n_{3}}{5}\right) {g}_1^{3}\,, \nonumber \\
		\beta^{(1)}(g_2) &=-\frac{19}{6} g_2^{3}+\left(  \frac{n_2}{2} + \frac{n_{\tilde{2}}}{2} + 2 n_3 \right) g_2^{3}\,, \nonumber \\
		\beta^{(1)}(g_3) &= -7 g_3^{3} +\left(  \frac{n_1}{6} + \frac{ n_{\tilde{1}}}{6} + \frac{n_2}{3} + \frac{n_{\tilde{2}}}{3} + \frac{n_{3}}{2} \right) g_3^{3}\,,
  \label{RGE1Loop}
\end{align}
where, $n_1$, $n_{\tilde{1}}$, $n_2$, $n_{\tilde{2}}$ and $n_3$ denote the number of copies, i.e.~generations, of $\Phi_{1}$, $\Phi_{\tilde{1}}$, $\Phi_{2}$, $\Phi_{\tilde{2}}$ and $\Phi_{3}$, respectively. The first term is the well-known SM term, and we have used the $SU(5)$ GUT normalization ${g}_1 = \sqrt{\frac{5}{3}}\, \tilde{g}_1$ w.r.t.~the values given in Table~\ref{SMfields}. The one-loop expressions for the $\beta$-function for the Yukawa couplings and the corresponding threshold correction are provided in Appendixes~\hyperlink{AppRGEYuk}{C} and \hyperlink{AppThreshold}{D}, respectively. The two-loop $\beta$-function coefficients for the gauge couplings are given in Appendix~\hyperlink{AppRGEGauge}{B}. 

\begin{figure}
    \centering
    \includegraphics[scale=0.6]{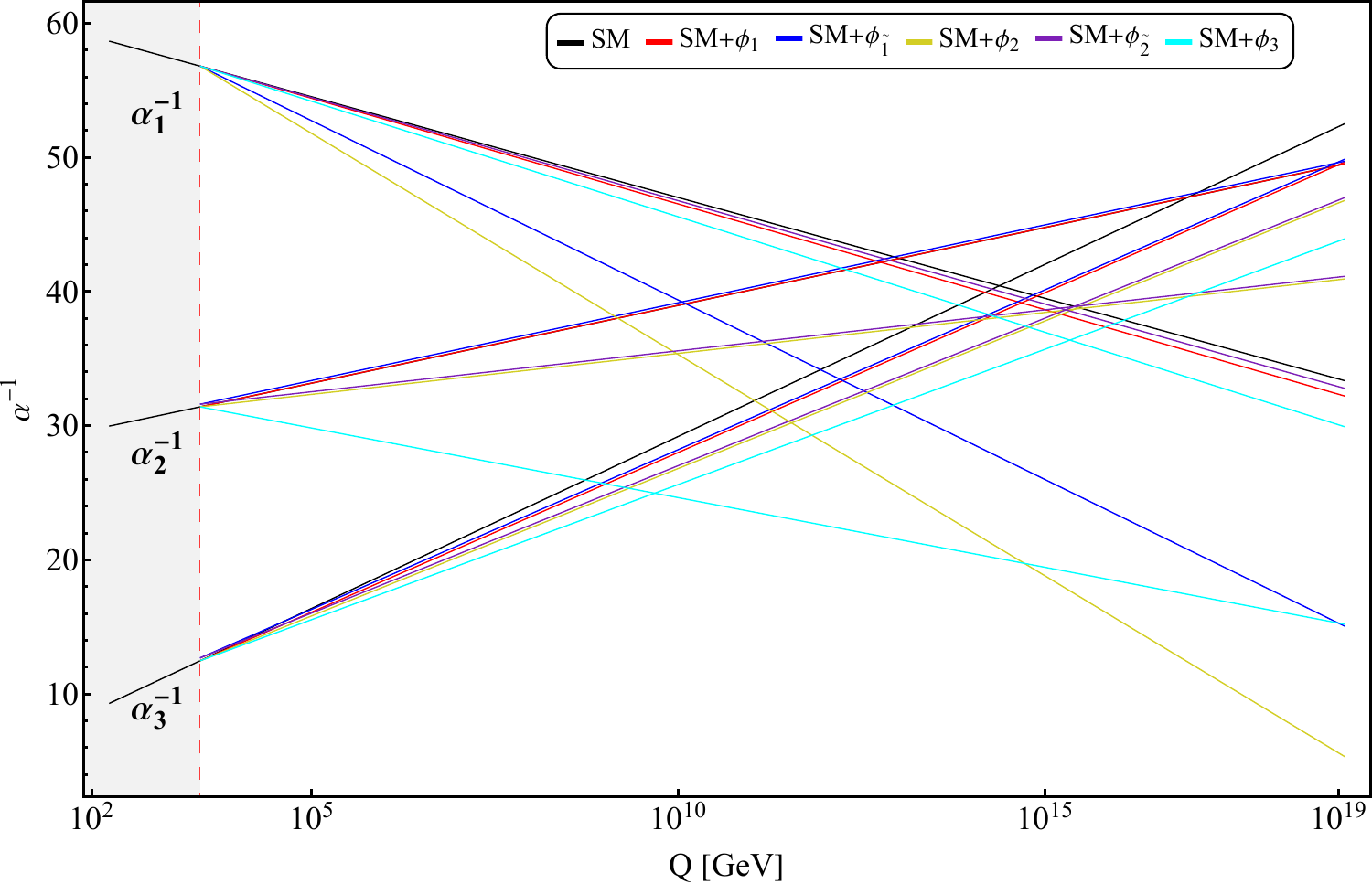}
    \caption{Renormalization group evolution of the gauge couplings $\left(\alpha_i = {g^2_i}/{4 \pi} \right)$ in SM and its extensions with the different LQ representations. In order to better visualize the effect of adding LQ, we considered  three generations of each LQ of mass 3$\,$TeV. For simplicity, we assumed all Yukawa and Higgs couplings involving LQs to be zero (at the LQ scale). }
    \label{MergedPlot3Gen}
\end{figure}

With these results at hand, we can next study the phenomenology of some selected cases. Here, we focus primarily on the evolution of the gauge and Higgs Yukawa couplings from the electroweak scale to the GUT scale.\footnote{Note that without a full specification of the GUT theory and its breaking sector, we cannot calculate the one-loop threshold corrections at the GUT scale which would be necessary to obtain a scheme and scale independent result. However, as no large logarithms are involved here, and the gauge coupling is already significantly smaller than the strong coupling at the EW scale, we assume these effects as uncertainties in the calculation.} As the starting point for the evolution, we take the following initial values of the SM parameters at the top scale, above the top threshold (i.e.~with 6 active flavours) in the $\overline{\text{\small MS}}$-scheme~\cite{Martin:2019lqd,Chetyrkin:2000yt,Herren:2017osy,Kniehl:2016enc,Huang:2020hdv,Degrassi:2014sxa,Degrassi:2012ry,Cheng:1973nv,Arason:1991ic,Ford:1992mv,Luo:2002ey,Tarasov:1982plg,Mihaila:2012fm,Bednyakov:2012en,Bednyakov:2013eba,vanRitbergen:1997va,Czakon:2004bu,Baikov:2016tgj,Luthe:2017ttc,Herzog:2017ohr,Jegerlehner:2003py,Marquard:2015qpa,Marquard:2016dcn,Martin:2016xsp,Gray:1990yh,Martin:2014cxa,Martin:2015rea,CMS:2021jnp,CMS:2015lbj,CMS:2018fks,Chetyrkin:2009fv,Beneke:2014pta,Narison:2019tym,BESIII:2014srs,ARGUS:1992chv}
\begin{align}
& g_1 = 0.3585 \times \sqrt{\frac{5}{3}}, \hspace{0.3cm}
g_2 = 0.648, \hspace{0.3cm}
g_3 = 1.16, \hspace{0.3cm}
\nonumber \\ &
y_t = 0.935, \hspace{0.3cm}
y_b  = 0.015, \hspace{0.3cm}
y_\tau = 0.01, \hspace{0.3cm}
\lambda = 0.126 \,.
\end{align}
We evolve the couplings using the $\beta$-functions of the SM up to the LQ scale $m_x$ (with $x=1,\tilde 1, 2,\tilde 2,3$). Note that here we neglected the tiny Yukawa couplings of the first two generations. 

After evolving the couplings to the LQ scale, we include the one-loop threshold corrections determined by comparing the theory with SLQs, to the one without them (i.e.~the SM at the LQ scale). This gives a shift to the SM fermion Yukawa couplings depending mainly on the initial values (at the LQ scale) of the LQ Yukawa couplings.\footnote{Note that for our SLQ models, the gauge couplings do not receive threshold corrections at one-loop in the $\overline{\text{\small MS}}$-scheme.} We then run all couplings, the SM ones as well as the ones including LQs, using the $\beta$-functions of the full model from the LQ scale to the high scale, for which we take as an upper limit the Planck scale $(\approx 10^{19})$ where gravitational effects would become important. 

\begin{figure}
    \centering
    \includegraphics[scale=0.295]{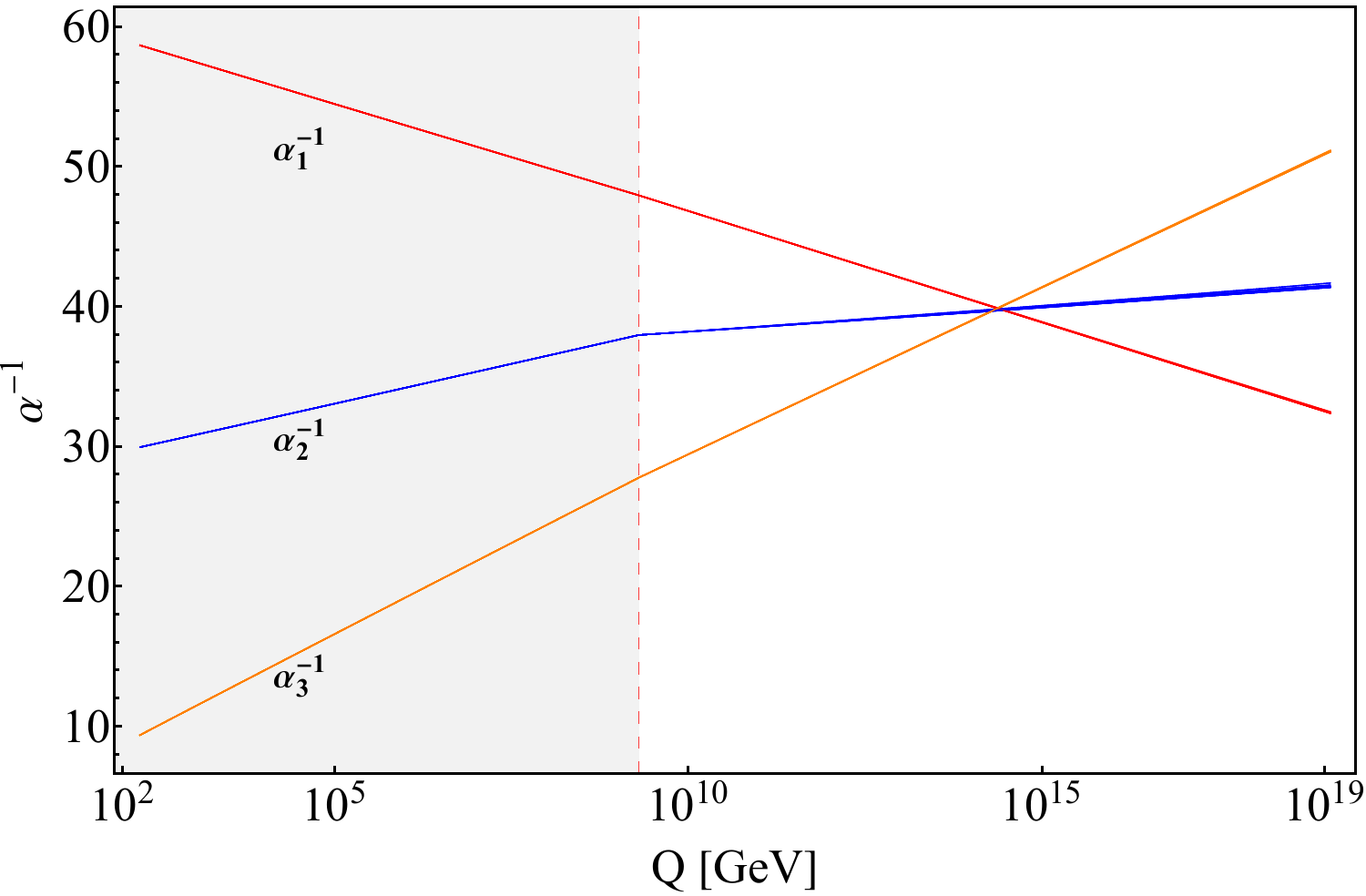}
        \includegraphics[scale=0.295]{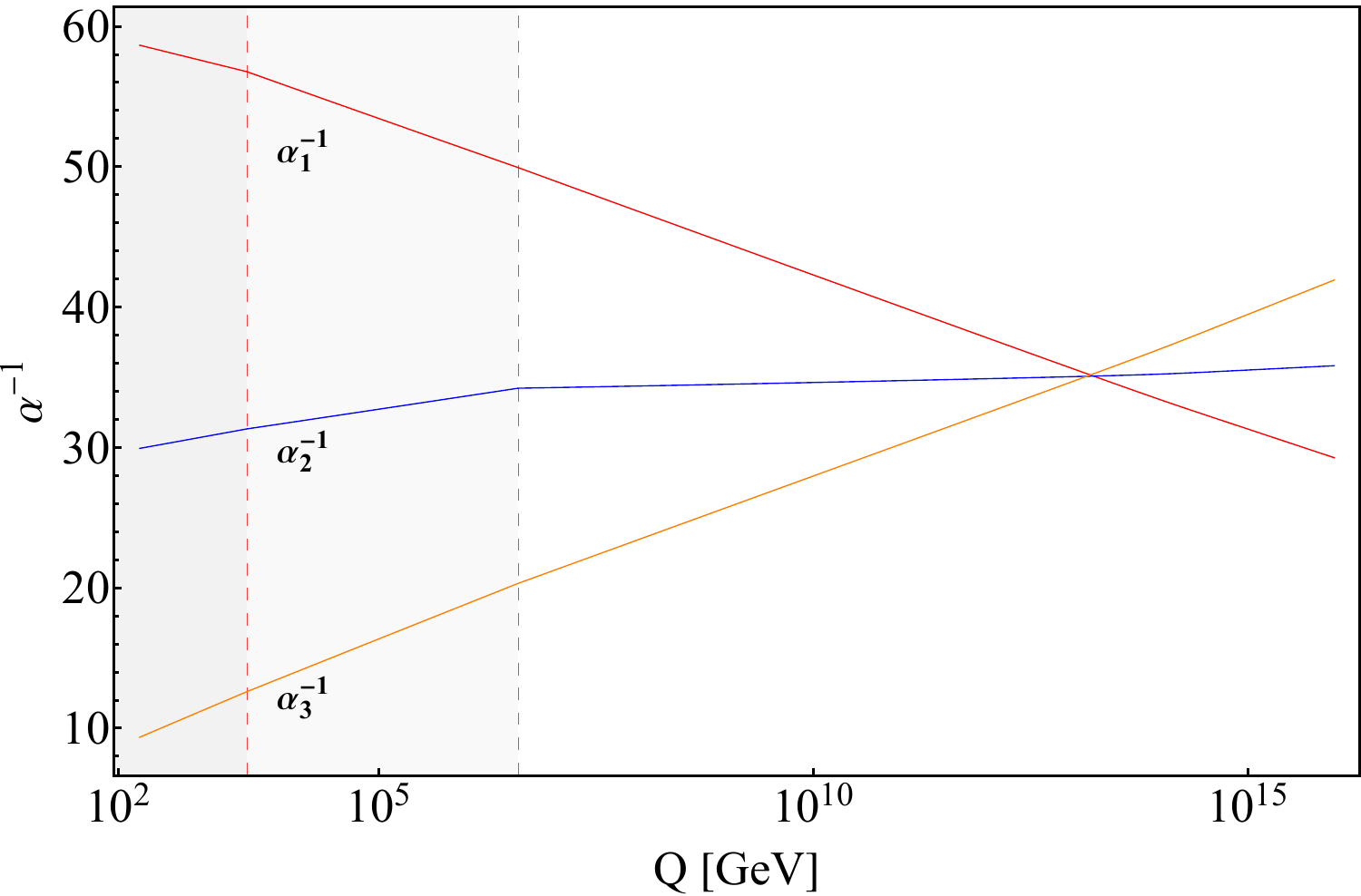}
\caption{Left: Two-loop renormalization group evolution of gauge couplings for SM$+ \Phi_{3}$. We observe gauge unification at around $10^{14}$\,GeV when we set the LQ scale to $m_{3} \approx 10^{6}$\,TeV. We vary the initial value of the three diagonal components of $Y^{LL}_{3}$ between $-0.6$ to $0.6$ (keeping the initial value of all other couplings to zero) to show that their effect on running is minimal. Right: Two-loop renormalization group running of gauge couplings for SM$+\Phi_{2} + \Phi_{3}$. We observe gauge unification for $m_2$= 3$\,$TeV and $m_3$= $4\times 10^{3}$\,TeV. 
%The initial value of the $(3,3)$ components of $Y^{RL}_2$ and $Y^{LR}_2$ are set to $0.1$ and $-0.1$, respectively. The remaining LQ Yukawa and quartic couplings' initial values are zero.
}
\label{GaugeCouplings3} 
\end{figure}

As a first step, we illustrate the running of the gauge coupling as a function of the scale $Q$ in Fig.~\ref{MergedPlot3Gen} for the five different LQ representations for a fixed LQ scale of $3\,$TeV, which is compatible with direct LHC searches ~\cite{ATLAS:2023prb,ATLAS:2023vxj,ATLAS:2023kek,ATLAS:2023uox,ATLAS:2022wcu,ATLAS:2021oiz,ATLAS:2020xov,ATLAS:2020dsk,ATLAS:2019qpq,ATLAS:2019ebv,CMS:2022goy,CMS:2020wzx,CMS:2018yiq,CMS:2018ncu,CMS:2018iye,CMS:2018oaj,CMS:2018lab,CMS:2018txo,CMS:2018qqq,CMS:2018svy}. To enhance and highlight the impact of adding the LQs, we considered three generations of the same LQ representation.

\begin{figure}
    \centering
    \includegraphics[scale=0.5]{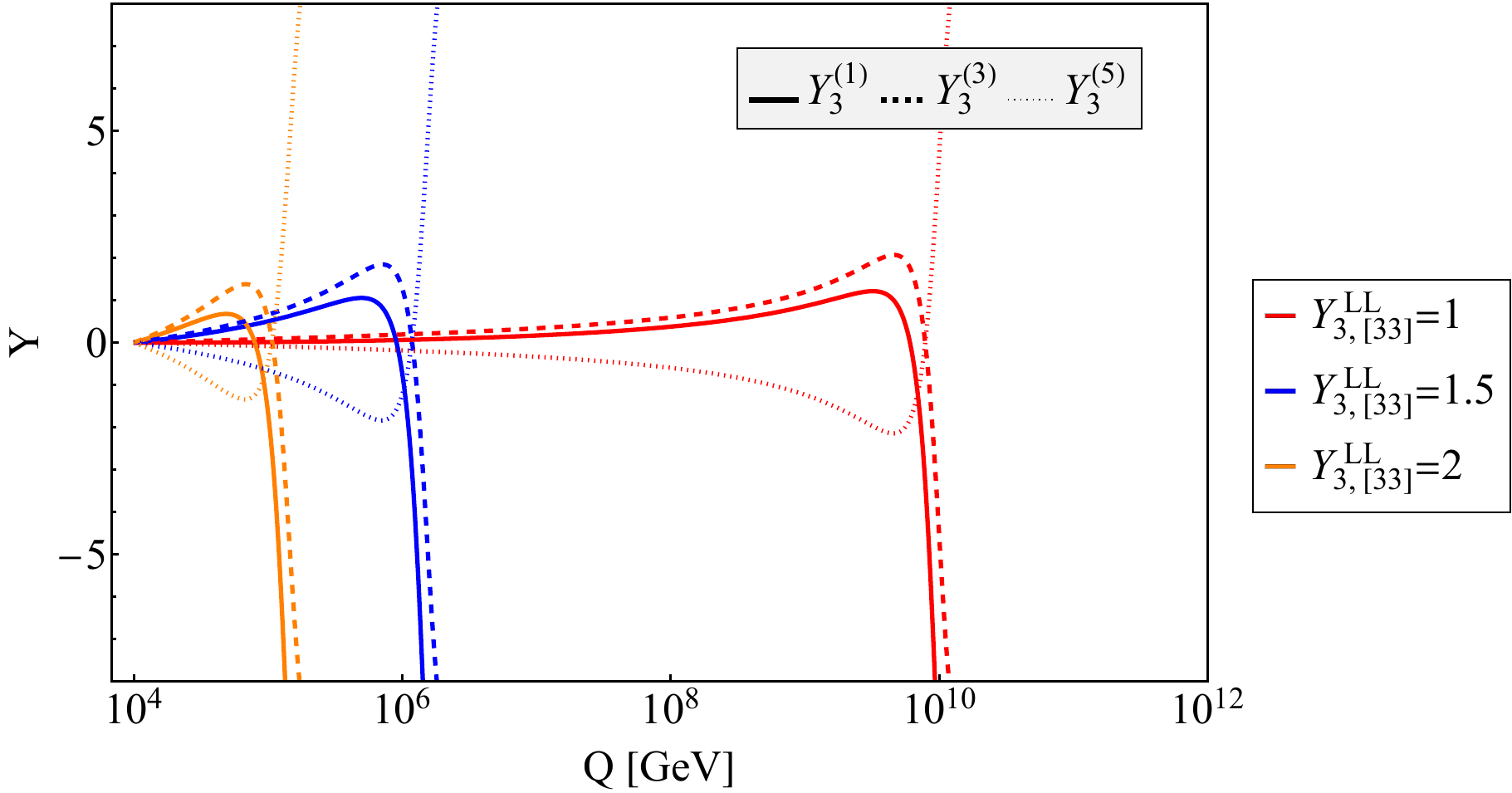}
    \caption{Two-loop renormalization group running of the quartic couplings $Y_3^{(1)}$ (solid), $Y_3^{(3)}$ (dashed) and $Y_3^{(5)}$ (dotted) in SM+$\Phi_3$ for different initial values at $m_3 \approx 10^4$ TeV of the $[3,3]$ component of the LQ Yukawa $Y^{LL}_3$ while assuming the other components to be small (as suggested by the $B$ anomalies). We observe that for larger (absolute) initial values of the LQ Yukawa, the quartic couplings become non-perturbative at lower scales. }
    \label{QuarticLandau}
\end{figure}

{We observe that no single generation LQ representation at the TeV scale can lead to gauge coupling unification. One can understand this behaviour of the gauge couplings from Eq.~\eqref{RGE1Loop}. Since $\beta^{(1)}(g_i)= a_i g^3_i$, where $a_i$ is a constant, $\alpha^{-1} = 4 \pi g^{-2}_i$ and the slope of the curve is $\alpha^{-1}_i ( \log (\mu))$ is $-8 \pi a_i$. We also see that w.r.t.~the SM curve (black) in Fig.~\ref{MergedPlot3Gen}, gauge coupling unification is improved if the slope of $\alpha^{-1}_1$ increases (i.e $a_1$ decreases) or if the slope of $\alpha^{-1}_2$ and $\alpha^{-1}_3$ decreases (i.e $a_2$ and $a_3$ increases). For $\Phi_1$ and $\Phi_{\bar{1}}$, $a_2$ remains unchanged but $a_1$ and $a_3$ increase. Therefore, $\Phi_1$ and $\Phi_{\bar{1}}$ will worsen the unification of gauge couplings. However, for the other three SLQs, both $a_2$ and $a_3$ increase, which improves the unification, despite $a_1$ also increasing. Specifically, $\Phi_{\bar{2}}$ and $\Phi_{3}$ improve the unification most since the increase in $a_1$ is much smaller. While the effect for $\Phi_{\bar{2}}$ is too weak, $\Phi_3$ can lead to gauge coupling unification at $\approx10^{14}\,$GeV for an LQ mass of around $10^6\,$TeV, as shown in Fig.~\ref{GaugeCouplings3} (left).\footnote{Such a low unification scale would conflict with the limits from proton decay~\cite{Super-Kamiokande:2020wjk}, at least for a standard $SU(5)$ GUT.}  Furthermore, gauge coupling unification is possible if we added three generations of $\Phi_{\bar{2}}$ as shown in Fig.~\ref{MergedPlot3Gen}.

Because the LQ Yukawa, Higgs and self-couplings have a minimal effect on the running of gauge couplings, we take their initial value to be zero at the LQ scale. In Fig.~\ref{GaugeCouplings3} (left), we also show the impact of including the LQ Yukawa couplings on SM fermions (which are free parameters). One can see that for the gauge coupling evolution, their effect is suppressed since they only enter at the two-loop level. In fact, we scanned over the initial values of all the three diagonal elements of $Y^{LL}_{3}$  between $-0.6$ and $0.6$ (while keeping the initial value of other couplings to zero), leading to a slight thickening of the curves. 

However, for larger values of the LQ Yukawa couplings non-perturbative values for some couplings, in particular quartic scalar couplings, can be induced. If one requires that this does not happen below a given scale, e.g.~the GUT scale, one can set an upper limit on the Yukawa couplings. We show the illustrative example of $\Phi_3$ in Fig.~\ref{QuarticLandau} where we plot three different quartic couplings for three large initial values (at $m_3 \approx 10^4$ TeV) of the $[3,3]$ component of $Y^{LL}_{3}$. One can see that the higher the initial value of the LQ Yukawa coupling, the lower the energy at which the quartic couplings reach non-perturbative values. Constraints on certain LQ Yukawa couplings were studied in~\cite{Bandyopadhyay:2021kue}.}

\begin{figure}
    \centering
    \includegraphics[scale=0.56]{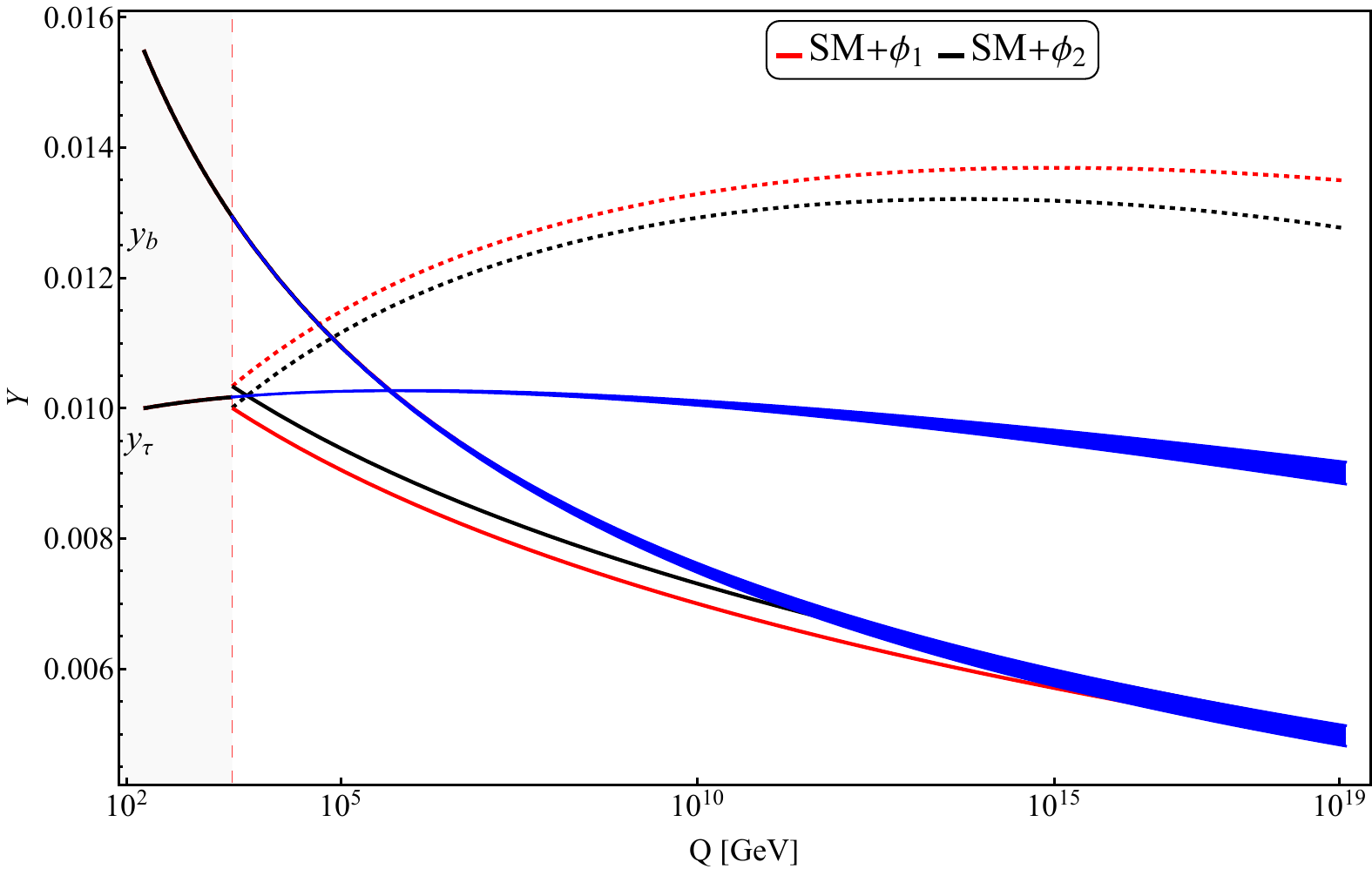}
    \caption{Two-loop renormalization group running of bottom and tau Yukawa couplings in SM+$\Phi_1$ and SM+$\Phi_2$ for a single-generation SLQ of mass 3\,TeV. The dotted line denotes the tau-Yukawa running when we set the initial value of the $(3,3)$ components of $Y^{RR}_1$, $Y^{LL}_1$, $Y^{RL}_2$ and $Y^{LR}_2$ to $0.1$. For solid lines, we reverse the sign of the $(3,3)$ components of $Y^{LL}_1$ and $Y^{LR}_2$. The blue bands show the variation of running in SM+$\Phi_i$, for $i=1,\tilde{1},2, \tilde{2} , 3$ for bottom Yukawa and $i=\tilde{1}, \tilde{2}, 3$ for tau Yukawa, while varying the $(3,3)$ components of LQ Yukawa couplings.}
    \label{MergedYukGen1}
\end{figure}

We now illustrate the RGE of bottom and tau Yukawa couplings in Fig.~\ref{MergedYukGen1}. While in the SM, bottom-tau Yukawa coupling unification already happens around $10^6$\,GeV, in LQ models, all representations have a small impact on the RGE of bottom Yukawa. Furthermore, the same is true for the tau one, except for $\Phi_1$ and $\Phi_2$. For them, the one-loop threshold corrections to the tau Yukawa coupling can be sizable and their running strongly changes (w.r.t.~the SM).\footnote{In this case, also a chirally enhanced effect in the anomalous magnetic moment of the $\tau$ lepton is generated~\cite{Crivellin:2021spu}. {This means that the chirality flips necessary to generate the anomalous magnetic moment does not originate from the rather small tau Yukawa coupling in the SM, but rather from a large coupling of new particles to the Higgs boson~\cite{Crivellin:2022wzw}.}} The running of the tau Yukawa coupling of SM+$\Phi_1$ and SM+$\Phi_2$ are shown in Fig.~\ref{MergedYukGen1}. The tau threshold correction strongly depends on our choice of the initial value of $(3,3)$ component of $Y^{RR}_{1}$, $Y^{LL}_{1}$, $Y^{RL}_{2}$ and $Y^{LR}_{2}$ which is set to $0.105$ for dotted lines and the sign of $Y^{LL}_{1}$ and $Y^{LR}_{2}$ reversed for the solid lines. The initial value of the remaining LQ couplings was set to zero. In fact, as can be seen in Fig.~\ref{MergedYukGen1}, the trajectories all lie within the quite narrow blue band except for, $\Phi_1$ and $\Phi_2$. Here a very strong impact on the evaluation of the tau Yukawa is possible due to the chiral enhancement. 

\begin{figure}
    \centering
    \includegraphics[scale=0.28]{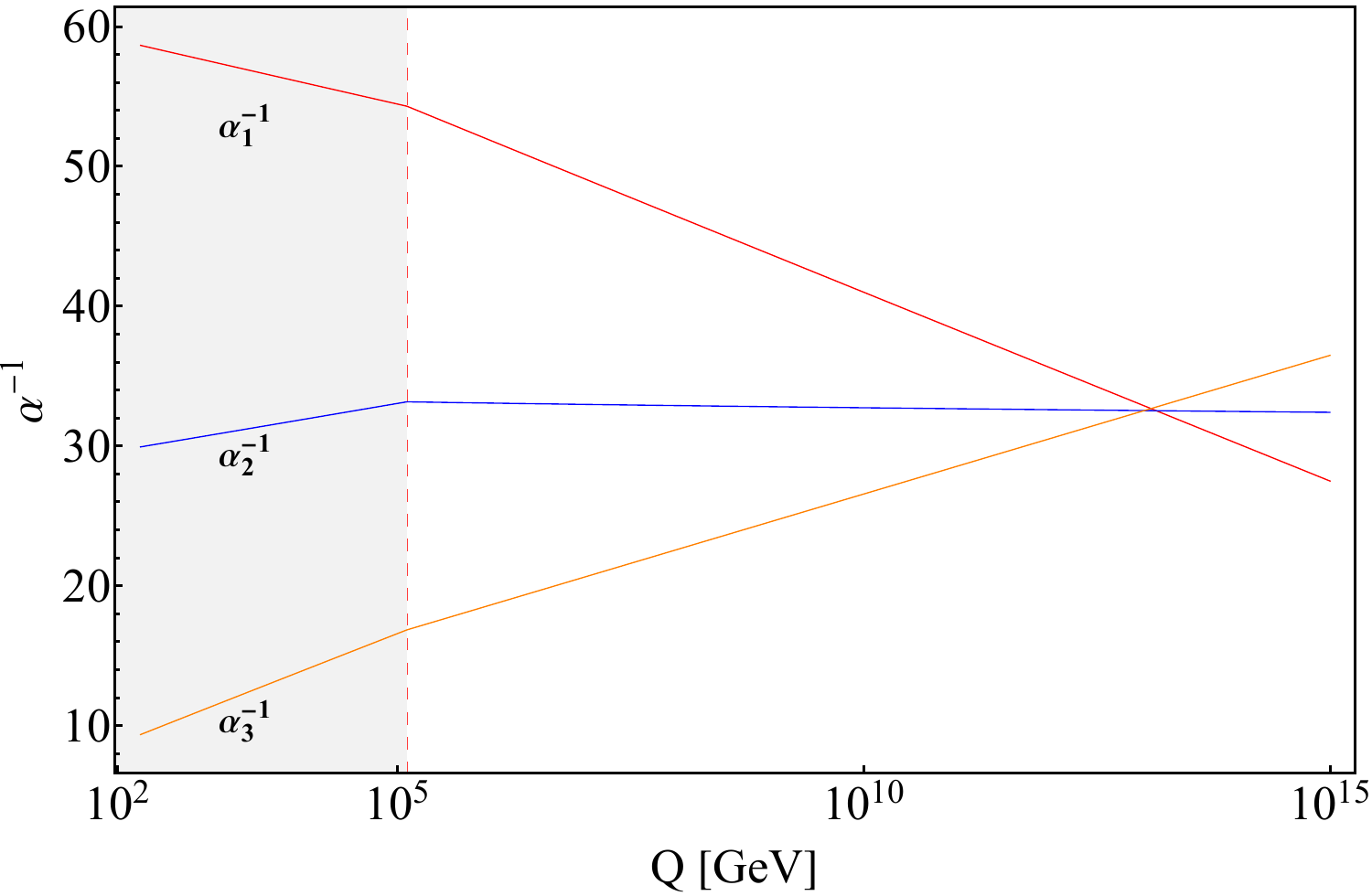}
     \includegraphics[scale=0.29]{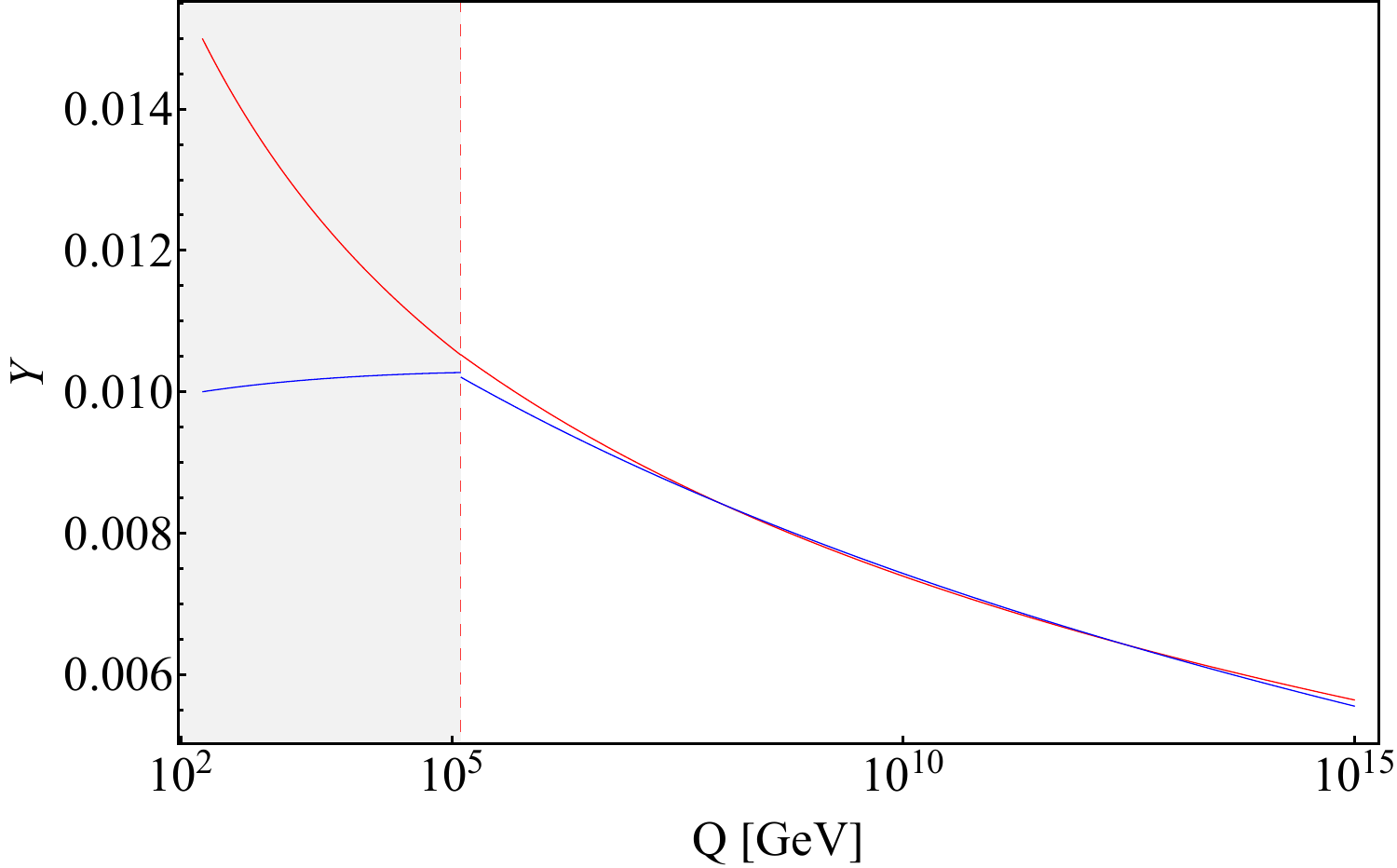}
\caption{Two-loop renormalization group running of gauge couplings (left), and bottom and tau Yukawa couplings (right) for ${\rm SM} +  \Phi_{1} +  \Phi_{\tilde{1}} +  \Phi_{2} + \Phi_{\tilde{2}} + \Phi_{3}$. We observe gauge coupling unification and bottom-tau unification at around the same energy scale $\approx 10^{13}$\,GeV when we set the mass of all LQs to $\approx 125$\,TeV and take the non-zero initial values of LQ couplings given in Eq.~\eqref{FullInitial}.}
\label{GaugeUnificationFull}
\end{figure}

We next consider the case, motivated by $SU(5)$ GUTs~\cite{Becirevic:2018afm}, where SM is extended by $\Phi_2$ and $\Phi_3$. Here, we find that for $m_2\approx 3$\,TeV and $m_3\approx10^{3.5}$\,TeV, we can achieve gauge coupling unification at $ \approx 10^{13}\,$TeV as shown in Fig.~\ref{GaugeCouplings3} (right) as well as Yukawa coupling unification around the same energy scale. However, this scale is again naively in conflict with proton decay. Finally, we consider the case where we extend the SM with (a single representation) all five possible SLQs (see Table~\ref{SMfields}). In this case, we obtain gauge coupling unification at around $10^{13}\,$TeV for a common LQ scale of $125\,$TeV as shown in Fig.~\ref{GaugeUnificationFull} (top). We also show in Fig.~\ref{GaugeUnificationFull} (bottom) that bottom-tau Yukawa coupling unification is possible nearly at the same energy scale for the following initial values of the $(3,3)$ component of the following  LQ Yukawa couplings at the LQ scale
\begin{align}\label{FullInitial}
    Y^{RR}_{1} {}_{[3,3]}= 0.1 \, , \, Y^{LL}_{1} {}_{[3,3]}= -0.1 \, , \,  Y^{RL}_{1} {}_{[3,3]}= -0.04 \, , \,  Y^{LR}_{1} {}_{[3,3]}= 0.1 \, , \, Y^{LL}_{3} {}_{[3,3]}= 0.45\,,
\end{align}
while taking the initial value of other LQ couplings to be zero.

\section{Conclusions}\label{sec4}

LQs are well-motivated extensions of the SM. They arise in composite or extra-dimensional setups and, most importantly, are predicted by GUTs. Furthermore, they have been under intensified investigation in the last years as they are potential candidates for describing several tensions between SM measurements and experiments.  

In this article, we computed the two-loop renormalization group evolution as well as the one-loop threshold corrections, for all parameters within generic SLQ models. This includes gauge couplings, SM Yukawa couplings, LQ Yukawa couplings (to quarks and leptons), Higgs and LQ (self-)interactions. The appendices collect the full SLQ Lagrangian, two-loop $\beta$-functions of the gauge couplings, one-loop threshold corrections and one-loop $\beta$-functions for the SM Yukawa couplings. The full analytic expressions, together with the necessary model files, can be obtained from \cite{GitHubSLQ}.

In our phenomenological analysis, we considered on the case in which one or more SLQs are light remnants of a symmetry-breaking sector of a GUT. In this setup, we focused on the renormalisation group evolution of the gauge and the Yukawa couplings, examining if unification can be achieved. Several simple scenarios were studied: 1) If one adds one of the 5 possible LQ representations to the SM, only $\Phi_3$ (with a mass around $10^6\,$TeV) can lead to gauge coupling unification at $\approx 10^{15}\,$GeV. 2) Extending the SM by all 5 possible LQ representations with a common mass scale $\approx 150\,$TeV, unification at $\approx 10^{14}\,$GeV is achieved. 3) Adding the GUT-motivated LQs $\Phi_2$ and  $\Phi_3$ to the SM, with masses of $m_2$= 3$\,$TeV and $m_3$= $4\times 10^{3}$\,TeV, respectively, gauge coupling unification occurs at $\approx 10^{13}\,$GeV. 4) Concerning the bottom-tau Yukawa unification, which happens at far too low scales in the SM, only the LQs $\Phi_1$ and $\Phi_2$ can lead to a sizable modification, due to possible chiral enhancement. Therefore, by choosing the initial value of LQ Yukawa couplings properly, one can always achieve bottom-tau unification at any desired (high) scale.

\section*{Acknowledgements}

We would like to thank Lohan Sartore for help in using \texttt{PyR@TE} and Anders E. Thomsen for help in using \texttt{RGBeta} and \texttt{Matchete}. We thank Michael Spira for useful comments on the draft. The work of A.C.~is supported by a professorship grant from the Swiss National Science Foundation (No.\ PP00P21\_76884).

\appendix

\hypertarget{AppLQLag}{\section{Lagrangian terms involving SLQs}\label{AppLQ}}
This appendix collects the terms in the SLQ Lagrangian in Eq. \eqref{LLQ} following the convention in \cite{Crivellin:2021ejk}. We begin with the terms $\mathcal{L}_{1}$, $\mathcal{L}_{{\tilde{1}}}$, $\mathcal{L}_{2}$, $\mathcal{L}_{{\tilde{2}}}$ and $\mathcal{L}_{3}$ which involve only single LQ interactions that include kinetic terms, mass terms, Yukawa interaction with SM fermions, quartic interaction with Higgs field and self-quartic interactions.
\begin{align} \label{L1}
\mathcal{L}_1 = & \left( D_\mu \Phi_1 \right)^{\dagger}D^\mu \Phi_1 - m_{1}^2 \Phi_1^{\dagger}\Phi_1 - Y_{1} \left(H^{\dagger}H\right) \Phi_1^{\dagger}\Phi_1   + \bigg[ Y_{1, ij}^{RR}\, \tilde{u}^{c}_i\ell_{j}\Phi_{1}^{\dagger}  +Y_{1, ij}^{LL}\, \left(\tilde{Q}_{i}^{c \intercal} i\sigma_{2}L_{j}\right)\Phi_{1}^{\dagger}
\nonumber \\ &
+Y_{1, ij}^{Q, LL}\, \left(\tilde{Q}^{c \intercal}_{i, c_1} i\sigma_2 Q_{j, c_2}\right) \Phi_{1, c_3} \epsilon^{c_1 c_2 c_3} + Y_{1, ij}^{Q, RR} \tilde{u}^{c}_{i, c1} d_{j, c_2} \Phi_{1, c_3} \epsilon^{c_1 c_2 c_3} +\text{h.c.} \bigg] \nonumber \\ & +\frac{1}{2} Y^{(1)}_{1} \left( \Phi_{1, c_1}^\dagger \Phi_{1, c_1} \right)\left(\Phi_{1, c_2}^\dagger \Phi_{1, c_2}\right)
\end{align}
\begin{align} \label{L1B}
\mathcal{L}_{\tilde{1}}= & \left( D_\mu \Phi_{\tilde{1}} \right)^{\dagger}D^\mu \Phi_{\tilde{1}} - m_{\tilde{1}}^2 \Phi_{\tilde{1}}^{\dagger}\Phi_{\tilde{1}} - Y_{\tilde{1}} \left(H^{\dagger}H\right) \Phi_{\tilde{1}}^{\dagger}\Phi_{\tilde{1}} + \bigg[ Y^{RR}_{\tilde{1}, ij}\, \tilde{d}^{  c}_{i}\ell_{j}\Phi_{\tilde{1}}^{\dagger}  
\nonumber \\ & + Y^{Q, RR}_{\tilde{1}, ij}\, \tilde{u}^{  c}_{i, c_1} u_{j, c_2}\Phi_{\tilde{1}, c_3} \epsilon^{c_1 c_2 c_3} +\text{h.c.} \bigg]
 +\frac{1}{2} Y^{(1)}_{\tilde{1}} \left( \Phi_{\tilde{1}, c_1}^\dagger \Phi_{\tilde{1}, c_1} \right)\left(\Phi_{\tilde{1}, c_2}^\dagger \Phi_{\tilde{1}, c_2}\right)
\end{align}
\begin{align} \label{L2}
\mathcal{L}_2= & \left( D_\mu \Phi_2 \right)^{\dagger}D^\mu \Phi_2 - m_{2}^2 \Phi_2^{\dagger}\Phi_2 - Y_{2} \left(H^{\dagger}H\right) \Phi_2^{\dagger}\Phi_2 +\bigg[ Y_{2, ij}^{RL}\, \left(\Phi_{2}^\intercal \tilde{u}_i i\sigma_2 L_{j} \right)+Y_{2, ij}^{LR}\, \left(\tilde{Q}_i \ell_j\Phi_{2}\right) \nonumber \\ & +\text{h.c.} \bigg]+\frac{1}{2} Y^{(1)}_{2} \left( \Phi_{2, c_1}^\dagger \Phi_{2, c_1} \right)\left(\Phi_{2, c_2}^\dagger \Phi_{2, c_2}\right) +\frac{1}{2} Y^{(3)}_{2} \left( \Phi_{2, c_1}^\dagger \Phi_{2, c_2} \right)\left(\Phi_{2, c_2}^\dagger \Phi_{2, c_1}\right) \nonumber \\ &
- Y_{22} \big(H^\intercal i\sigma_{2} \Phi_2\big)^\dagger \big(H^\intercal i\sigma_{2} \Phi_2\big)
\end{align}
\begin{align} \label{L2B}
\mathcal{L}_{\tilde{2}}= & \left( D_\mu \Phi_{\tilde{2}} \right)^{\dagger}D^\mu \Phi_{\tilde{2}} - m_{{\tilde{2}}}^2 \Phi_{\tilde{2}}^{\dagger}\Phi_{\tilde{2}} - Y_{\tilde{2}} \left(H^{\dagger}H\right) \Phi_{\tilde{2}}^{\dagger}\Phi_{\tilde{2}} +\bigg[ Y_{\tilde{2}, ij}^{RL}\, \left(\Phi_{\tilde{2}}^\intercal \tilde{d}_{i} i\sigma_2 L_{j}\right)+\text{h.c.} \bigg] 
\nonumber \\ &
 +\frac{1}{2} Y^{(1)}_{\tilde{2}} \left( \Phi_{\tilde{2}, c_1}^\dagger \Phi_{\tilde{2}, c_1} \right)\left(\Phi_{\tilde{2}, c_2}^\dagger \Phi_{\tilde{2}, c_2}\right)+\frac{1}{2} Y^{(3)}_{\tilde{2}} \left( \Phi_{\tilde{2}, c_1}^\dagger \Phi_{\tilde{2}, c_2} \right)\left(\Phi_{\tilde{2}, c_2}^\dagger \Phi_{\tilde{2}, c_1}\right)
 \nonumber \\ &
 - Y_{\tilde{2}\tilde{2}} \big(H^\intercal i\sigma_{2} \Phi_{\tilde{2}} \big)^\dagger \big(H^\intercal i\sigma_{2} \Phi_{\tilde{2}} \big)
\end{align}
\begin{align} \label{L3}
\mathcal{L}_3=& \left( D_\mu \Phi_3 \right)^{\dagger}D^\mu \Phi_3 - m_{3}^2 \Phi_3^{\dagger}\Phi_3 - Y_{3} \left(H^{\dagger}H\right) \Phi_3^{\dagger}\Phi_3 +\bigg[ Y_{3, ij}^{LL}\, \left(\tilde{Q}^{c \intercal}_{i} i\sigma_{2}\left(\sigma\cdot\Phi_{3}\right)^{\dagger}L_{j}\right) 
\nonumber \\ &
 + Y^{Q, LL}_{3, ij}\, \left(\tilde{Q}^{ c \intercal}_{i, c_1} i\sigma_2 \left(\sigma \cdot \Phi_{3, c_3}\right)Q_{j, c_2}\right) \epsilon^{c_1 c_2 c_3} +\text{h.c.}\bigg]  +\frac{1}{2} Y^{(1)}_{3} \left( \Phi_{3, c_1}^\dagger \Phi_{3, c_1} \right)\left(\Phi_{3, c_2}^\dagger \Phi_{3, c_2}\right) 
 \nonumber \\ & 
 +\frac{1}{2} Y^{(3)}_{3} \left( \Phi_{3, c_1}^\dagger \Phi_{3, c_2} \right)\left(\Phi_{3, c_2}^\dagger \Phi_{3, c_1}\right) - i Y_{33} \epsilon^{IJK} \left(H^\dagger \sigma^I H \right) \Phi_3^{J\dagger} \Phi_3^K 
 \nonumber \\ &+ \frac{1}{2} Y^{(5)}_{3} \big( \Phi_{3, c_1}^{I\dagger} \Phi_{3, c_1}^{J}  \Phi_{3, c_2}^{I\dagger} \Phi_{3, c_2}^{J}\big)
\end{align}

where we have omitted, at several places, color indices involving trivial contraction. For example, $Y_{1} \left(H^{\dagger}H\right) \Phi_1^{\dagger}\Phi_1 $ in full notation is $Y_{1} \left(H^{\dagger}H\right) \Phi_{1 \, c_1}^{\dagger}\Phi_{1 \, c_1}$.

The next set of terms $\mathcal{L}_{1\,{\tilde{1}}}$, $\mathcal{L}_{1\,{2}}$, $\mathcal{L}_{1\,{\tilde{2}}}$, $\mathcal{L}_{1\,{3}} $, $\mathcal{L}_{{\tilde{1}}\,{2}}$, $\mathcal{L}_{{\tilde{1}}\,{\tilde{2}}}$, $\mathcal{L}_{{\tilde{1}}\,{3}}$, $\mathcal{L}_{{2}\,{\tilde{2}}}$, $\mathcal{L}_{{2}\,{3}}$ and $\mathcal{L}_{{\tilde{2}}\,{3}}$ contain interactions involving precisely two LQs.

\begin{align} \label{L11B}
\mathcal{L}_{1\,{\tilde{1}}} =  Y^{(1)}_{1 \tilde{1}} \left( \Phi_{1, c_1}^\dagger \Phi_{1, c_1} \right) \left( \Phi_{\tilde{1}, c_2}^\dagger \Phi_{\tilde{1}, c_2} \right) + Y^{(\prime{1})}_{1 \tilde{1}} \left( \Phi_{1, c_1}^\dagger \Phi_{1, c_2} \right) \left( \Phi_{\tilde{1}, c_2}^\dagger \Phi_{\tilde{1}, c_1} \right)
\end{align}

\begin{align} \label{L12}
\mathcal{L}_{1\,{2}} = Y^{(1)}_{1 {2}} \left( \Phi_{1, c_1}^\dagger \Phi_{1, c_1}\right) \left( \Phi_{{2}, c_2}^\dagger \Phi_{{2}, c_2} \right) + Y^{\prime(1)}_{1 {2}} \left( \Phi_{1, c_1}^\dagger \Phi_{1, c_2}\right) \left( \Phi_{{2}, c_2}^\dagger \Phi_{{2}, c_1} \right)
\end{align}

\begin{align} \label{L12B}
\mathcal{L}_{1\,{\tilde{2}}} = &  Y^{(1)}_{1 \tilde{2}} \left( \Phi_{1, c_1}^\dagger \Phi_{1, c_1}\right) \left( \Phi_{\tilde{2}, c_2}^\dagger \Phi_{\tilde{2}, c_2} \right) + Y^{\prime(1)}_{1 \tilde{2}} \left( \Phi_{1, c_1}^\dagger \Phi_{1, c_2}\right) \left( \Phi_{\tilde{2}, c_2}^\dagger \Phi_{\tilde{2}, c_1} \right)
 \nonumber \\ & 
 +\bigg[ A_{1\tilde{2}\tilde{2}} \ \Phi_{1, c_1} \left(\Phi_{\tilde{2}, c_2}^\intercal i\sigma_2 \Phi_{\tilde{2}, c_3}\right) \epsilon^{c_1 c_2 c_3} - A_{1\tilde{2}} \, \Phi_1 \left( \Phi_{\tilde{2}}^\dagger H \right) +\text{h.c.}\bigg]
\end{align}

\begin{align} \label{L13}
\mathcal{L}_{1\,{3}} = &  Y^{(1)}_{1 {3}} \left( \Phi_{1, c_1}^\dagger \Phi_{1, c_1}\right) \left( \Phi_{{3}, c_2}^\dagger \Phi_{{3}, c_2} \right) + Y^{\prime(1)}_{1 {3}} \left( \Phi_{1, c_1}^\dagger \Phi_{1, c_2}\right) \left( \Phi_{{3}, c_2}^\dagger \Phi_{{3}, c_1} \right) + \bigg[ Y_{13} \big(H^{\dagger}\left(\sigma\cdot\Phi_{3} \right)H \big)\Phi_{1}^{\dagger}
\nonumber \\ &
+ \frac{1}{2} Y_{1 3 1 3} \left(\Phi^\dagger_{1, c_1} \Phi^I_{3, c_1} \Phi^\dagger_{1, c_2} \Phi^I_{3, c_2} \right) + Y_{1 3 3 3} \left(\Phi^\dagger_{1, c_1} \Phi^I_{3, c_1} \Phi^{J \dagger}_{3, c_2} \Phi^K_{3, c_2} i\epsilon^{IJK} \right) +\text{h.c.}\bigg]
\end{align}

\begin{align} \label{L1B2}
\mathcal{L}_{{\tilde{1}}\,{2}} = Y^{(1)}_{\tilde{1} {2}} \left( \Phi_{\tilde{1}, c_1}^\dagger \Phi_{\tilde{1}, c_1}\right) \left( \Phi_{{2}, c_2}^\dagger \Phi_{{2}, c_2} \right) + Y^{\prime(1)}_{\tilde{1} {2}} \left( \Phi_{\tilde{1}, c_1}^\dagger \Phi_{\tilde{1}, c_2}\right) \left( \Phi_{{2}, c_2}^\dagger \Phi_{{2}, c_1} \right) 
\end{align}

\begin{align} \label{L1B2B}
\mathcal{L}_{{\tilde{1}}\,{\tilde{2}}} =  Y^{(1)}_{\tilde{1} \tilde{2}} \left( \Phi_{\tilde{1}, c_1}^\dagger \Phi_{\tilde{1}, c_1}\right) \left( \Phi_{\tilde{2}, c_2}^\dagger \Phi_{\tilde{2}, c_2} \right) + Y^{\prime(1)}_{\tilde{1} \tilde{2}} \left( \Phi_{\tilde{1}, c_1}^\dagger \Phi_{\tilde{1}, c_2}\right) \left( \Phi_{\tilde{2}, c_2}^\dagger \Phi_{\tilde{2}, c_1} \right)
\end{align}

\begin{align} \label{L1B3}
\mathcal{L}_{{\tilde{1}}\,{3}} = &  Y^{(1)}_{\tilde{1} {3}} \left( \Phi_{\tilde{1}, c_1}^\dagger \Phi_{\tilde{1}, c_1}\right) \left( \Phi_{{3}, c_2}^\dagger \Phi_{{3}, c_2} \right) + Y^{\prime(1)}_{\tilde{1} {3}} \left( \Phi_{\tilde{1}, c_1}^\dagger \Phi_{\tilde{1}, c_2}\right) \left( \Phi_{{3}, c_2}^\dagger \Phi_{{3}, c_1} \right)
\nonumber \\ &
+\bigg[Y_{\tilde{1}3}\big(H^\intercal i\sigma_{2}\left(\sigma\cdot\Phi_{3}\right)^\dagger H\big)\Phi_{\tilde{1}} + \text{h.c.}\bigg]
\end{align}

\begin{align} \label{L22B}
\mathcal{L}_{{2}\,{\tilde{2}}} = &  Y^{(1)}_{{2} \tilde{2}} \left( \Phi_{{2}, c_1}^\dagger \Phi_{{2}, c_1}\right) \left( \Phi_{\tilde{2}, c_2}^\dagger \Phi_{\tilde{2}, c_2} \right) + Y^{\prime(1)}_{{2} \tilde{2}} \left( \Phi_{{2}, c_1}^\dagger \Phi_{{2}, c_2}\right) \left( \Phi_{\tilde{2}, c_2}^\dagger \Phi_{\tilde{2}, c_1} \right) 
\nonumber \\ & 
+Y^{(3)}_{{2} \tilde{2}} \left( \Phi_{{2}, c_1}^\dagger \Phi_{\tilde{2}, c_1}\right) \left( \Phi_{\tilde{2}, c_2}^\dagger \Phi_{2, c_2} \right)
 + Y^{\prime(3)}_{{2} \tilde{2}} \left( \Phi_{{2}, c_1}^\dagger \Phi_{\tilde{2}, c_2}\right) \left( \Phi_{\tilde{2}, c_2}^\dagger \Phi_{2, c_1} \right)
\nonumber \\ & 
+ \bigg[ Y_{2\tilde{2}}\left(\Phi_{2}^\dagger H \right)\left(H^\intercal i\sigma_{2}\Phi_{\tilde{2}}\right) + \text{h.c.} \bigg]
\end{align}

\begin{align} \label{L23}
\mathcal{L}_{{2}\,{3}} = &  Y^{(1)}_{{2} {3}} \left( \Phi_{{2}, c_1}^\dagger \Phi_{{2}, c_1}\right) \left( \Phi_{{3}, c_2}^\dagger \Phi_{{3}, c_2} \right) + Y^{\prime(1)}_{{2} {3}} \left( \Phi_{{2}, c_1}^\dagger \Phi_{{2}, c_2}\right) \left( \Phi_{{3}, c_2}^\dagger \Phi_{{3}, c_1} \right)
\nonumber \\ &
+Y^{(3)}_{{2} {3}} \left( \Phi_{{2}, c_1}^\dagger \sigma^{I} \Phi_{{2}, c_1}\right) \left( \Phi_{{3}, c_2}^{J \dagger} i \epsilon^{ I J K} \Phi_{{3}, c_2}^{K} \right)
+Y^{\prime(3)}_{{2} {3}} \left( \Phi_{{2}, c_1}^\dagger \sigma^{I} \Phi_{{2}, c_2}\right) \left( \Phi_{{3}, c_2}^{J \dagger} i \epsilon^{ I J K} \Phi_{{3}, c_1}^{K} \right) 
\nonumber \\ & 
+ \bigg[ Y_{233} \ \left(H^\dagger \sigma^I \Phi_{2, c_1}\right)\left( \Phi^J_{3, c_2} i \epsilon^{IJK} \Phi^K_{3, c_3}\right) \epsilon^{c_1 c_2 c_3} +\text{h.c.}\bigg]
\end{align}

\begin{align} \label{L2B3}
\mathcal{L}_{{\tilde{2}}\,{3}}  = &  Y^{(1)}_{\tilde{2} {3}} \left( \Phi_{\tilde{2}, c_1}^\dagger \Phi_{\tilde{2}, c_1}\right) \left( \Phi_{{3}, c_2}^\dagger \Phi_{{3}, c_2} \right) + Y^{\prime(1)}_{\tilde{2} {3}} \left( \Phi_{\tilde{2}, c_1}^\dagger \Phi_{\tilde{2}, c_2}\right) \left( \Phi_{{3}, c_2}^\dagger \Phi_{{3}, c_1} \right)
\nonumber \\ &
  +Y^{\prime(3)}_{\tilde{2} {3}} \left( \Phi_{\tilde{2}, c_1}^\dagger \sigma^{I} \Phi_{\tilde{2}, c_2}\right) \left( \Phi_{{3}, c_2}^{J \dagger} i \epsilon^{ I J K} \Phi_{{3}, c_1}^{K} \right) 
  +Y^{(3)}_{\tilde{2} {3}} \left( \Phi_{\tilde{2}, c_1}^\dagger \sigma^{I} \Phi_{\tilde{2}, c_1}\right) \left( \Phi_{{3}, c_2}^{J \dagger} i \epsilon^{ I J K} \Phi_{{3}, c_2}^{K} \right)
\nonumber \\ & 
+\bigg[
Y_{\tilde{2}33} \ \left(\Phi_{\tilde{2}, c_1}^\intercal i \sigma_2 \sigma^I H\right) \left( \Phi^J_{3, c_2} i\epsilon^{IJK}\Phi^K_{3, c_3} \right) \epsilon^{c_1 c_2 c_3}
+ A_{\tilde{2} 3}\left(\Phi_{\tilde{2}}^\dagger \left(\sigma \cdot \Phi_3\right) H \right)+ \text{h.c.} \bigg]
\end{align}

The next set of terms $\mathcal{L}_{{\tilde{1}}\,{2}\,{\tilde{2}}}$, $\mathcal{L}_{{1}\,{\tilde{1}}\,{2}}$, $\mathcal{L}_{{1}\,{2}\,{3}}$, $\mathcal{L}_{{1}\,{\tilde{2}}\,{3}}$ and $\mathcal{L}_{{\tilde{1}}\,{2}\,{3}}$ contain interactions involving precisely three LQs.

\begin{align} \label{L1B22B}
\mathcal{L}_{{\tilde{1}}\,{2}\,{\tilde{2}}} =A_{\tilde{1}2\tilde{2}} \ \Phi_{\tilde{1}, c_1} \left(\Phi_{2, c_2}^\intercal i\sigma_2 \Phi_{\tilde{2}, c_3}\right) \epsilon^{c_1 c_2 c_3} + \text{h.c.}
\end{align}

\begin{align} \label{L11B2}
\mathcal{L}_{{1}\,{\tilde{1}}\,{2}} = Y_{1 \tilde{1}2} \ \Phi_{1, c_1} \Phi_{\tilde{1}, c_2} \left( \Phi_{2, c_3}^\intercal i \sigma_2 H \right) \epsilon^{c_1 c_2 c_3} + \text{h.c.}
\end{align}

\begin{align} \label{L123}
\mathcal{L}_{{1}\,{2}\,{3}} = & Y_{123} \ \Phi_{1, c_1} \left(H^\dagger \left( \sigma \cdot \Phi_{3, c_3} \right) \Phi_{2, c_2}\right) \epsilon^{c_1 c_2 c_3} + Y_{1223} \Phi^\dagger_{1, c_1}  \left(\Phi_{2, c_2}^\dagger \big(\sigma \cdot \Phi_{3, c_2} \big) \Phi_{2, c_1} \right) 
\nonumber \\ &
 + Y^\prime_{1223}  \Phi^\dagger_{1, c_1}  \left(\Phi_{2, c_2}^\dagger \big(\sigma \cdot \Phi_{3, c_1} \big) \Phi_{2, c_2} \right) + \text{h.c.}
\end{align}

\begin{align} \label{L12B3}
\mathcal{L}_{{1}\,{\tilde{2}}\,{3}} = & Y_{1\tilde{2}3} \ \Phi_{1, c_1}  \left(\Phi_{\tilde{2}, c_2}^\intercal i \sigma_2 \left( \sigma \cdot \Phi_{3, c_3} \right) H\right) \epsilon^{c_1 c_2 c_3} + Y_{1\tilde{2}\tilde{2}3} \Phi^\dagger_{1, c_1}  \left(\Phi_{\tilde{2}, c_2}^\dagger \big(\sigma \cdot \Phi_{3, c_2} \big) \Phi_{\tilde{2}, c_1} \right) 
\nonumber \\ &
 + Y^\prime_{1\tilde{2}\tilde{2}3}  \Phi^\dagger_{1, c_1}  \left(\Phi_{\tilde{2}, c_2}^\dagger \big(\sigma \cdot \Phi_{3, c_1} \big) \Phi_{\tilde{2}, c_2} \right) + \text{h.c.}
\end{align}

\begin{align} \label{L1B23}
\mathcal{L}_{{\tilde{1}}\,{2}\,{3}} = Y_{\tilde{1}23} \ \Phi_{\tilde{1}, c_1}  \left(\Phi_{2, c_2}^\intercal i \sigma_2 \left( \sigma \cdot \Phi_{3, c_3} \right) H\right) \epsilon^{c_1 c_2 c_3} + \text{h.c.}
\end{align}

Finally, the interaction terms $\mathcal{L}_{{\tilde{1}}\,{2}\,{\tilde{2}}\,{3}}$ and $\mathcal{L}_{{1}\,{\tilde{1}}\,{2}\,{\tilde{2}}}$ which involve four different  SLQs are as follows,

\begin{align} \label{L1B22B3}
\mathcal{L}_{{\tilde{1}}\,{2}\,{\tilde{2}}\,{3}} =  Y_{\tilde{1} \tilde{2} 2 3} \Phi_{\tilde{1}, c_1}^\dagger  \left(\Phi_{2, c_2}^\dagger \big(\sigma \cdot \Phi_{3, c_2} \big)\Phi_{\tilde{2}, c_1} \right) + Y^\prime_{\tilde{1} \tilde{2} 2 3} \Phi_{\tilde{1}, c_1}^\dagger  \left(\Phi_{2, c_2}^\dagger \big(\sigma \cdot \Phi_{3, c_1} \big)\Phi_{\tilde{2}, c_2} \right) + \text{h.c.}
\end{align}

\begin{align} \label{L11B22B}
\mathcal{L}_{{1}\,{\tilde{1}}\,{2}\,{\tilde{2}}} = Y_{1 \tilde{1} 2 \tilde{2}} \Phi^\dagger_{1, c_1} \Phi_{\tilde{1}, c_1} \left( \Phi_{\tilde{2}, c_2}^\dagger  \Phi_{2, c_2} \right) + Y^\prime_{1 \tilde{1} 2 \tilde{2}}  \Phi^\dagger_{1, c_1} \Phi_{\tilde{1}, c_2} \left( \Phi_{\tilde{2}, c_2}^\dagger  \Phi_{2, c_1} \right) + \text{h.c.}
\end{align}

\hypertarget{AppRGEGauge}{\section{Two-loop $\beta$-functions of gauge couplings}}
In this Appendix, we collect the two-loop $\beta$-functions of gauge  couplings of some of the SLQ models considered in this article.
\begin{itemize}
    \item  SM $ + \, \Phi_1$
    {\allowdisplaybreaks
\begin{align}
\begin{autobreak}
\beta^{(2)}(g_1) =
+ \frac{121}{30} g_1^{5}
+ \frac{27}{10} g_1^{3} g_2^{2}

+ \frac{148}{15} g_1^{3} g_3^{2}

-  \frac{17}{10} g_1^{3} \tr\left(Y_u^{\dagger} Y_u \right)

-  \frac{1}{2} g_1^{3} \tr\left(Y_d^{\dagger} Y_d \right)

-  \frac{3}{2} g_1^{3} \tr\left(Y_e^{\dagger} Y_e \right)

-  \frac{13}{5} g_1^{3} \tr\left({Y^{RR}_1}^{\dagger} Y^{RR}_1 \right)

-  g_1^{3} \tr\left({Y^{LL}_1}^{\dagger} Y^{LL}_1 \right)

-  \frac{4}{5} g_1^{3} \tr\left(Y^{Q,LL\,*}_1 Y^{Q,LL}_1 \right)

- 2 g_1^{3} \tr\left({Y^{Q,RR}_1}^{\dagger} Y^{Q,RR}_1 \right)
\end{autobreak}
\end{align}
\vspace{-0.5cm}
\begin{align}
\begin{autobreak}
\beta^{(2)}(g_2) =

+ \frac{9}{10} g_1^{2} g_2^{3}

+ \frac{35}{6} g_2^{5}

+ 12 g_2^{3} g_3^{2}

-  \frac{3}{2} g_2^{3} \tr\left(Y_u^{\dagger} Y_u \right)

-  \frac{3}{2} g_2^{3} \tr\left(Y_d^{\dagger} Y_d \right)

-  \frac{1}{2} g_2^{3} \tr\left(Y_e^{\dagger} Y_e \right)

- 3 g_2^{3} \tr\left({Y^{LL}_1}^{\dagger} Y^{LL}_1 \right)

- 12 g_2^{3} \tr\left(Y^{Q,LL\,*}_1 Y^{Q,LL}_1 \right)
\end{autobreak}
\end{align}
\vspace{-0.5cm}
\begin{align}
\begin{autobreak}
\beta^{(2)}(g_3) =

+ \frac{37}{30} g_1^{2} g_3^{3}

+ \frac{9}{2} g_2^{2} g_3^{3}

-  \frac{67}{3} g_3^{5}

- 2 g_3^{3} \tr\left(Y_u^{\dagger} Y_u \right)

- 2 g_3^{3} \tr\left(Y_d^{\dagger} Y_d \right)

-  \frac{1}{2} g_3^{3} \tr\left({Y^{RR}_1}^{\dagger} Y^{RR}_1 \right)

-  g_3^{3} \tr\left({Y^{LL}_1}^{\dagger} Y^{LL}_1 \right)

- 8 g_3^{3} \tr\left(Y^{Q,LL\,*}_1 Y^{Q,LL}_1 \right)

- 2 g_3^{3} \tr\left({Y^{Q,RR}_1}^{\dagger} Y^{Q,RR}_1 \right)
\end{autobreak}
\end{align}
}
\item {SM $ + \, \Phi_{\tilde{1}}$}
\begin{align}
\begin{autobreak}
\beta^{(2)}(g_1) =

+ \frac{529}{30} g_1^{5}

+ \frac{27}{10} g_1^{3} g_2^{2}

+ \frac{388}{15} g_1^{3} g_3^{2}

-  \frac{17}{10} g_1^{3} \tr\left(Y_u^{\dagger} Y_u \right)

-  \frac{1}{2} g_1^{3} \tr\left(Y_d^{\dagger} Y_d \right)

-  \frac{3}{2} g_1^{3} \tr\left(Y_e^{\dagger} Y_e \right)

- 2 g_1^{3} \tr\left({Y^{RR}_{{\tilde{1}}}}^{\dagger} Y^{RR}_{{\tilde{1}}} \right)

+ \frac{32}{5} g_1^{3} \tr\left(Y^{Q,RR\,*}_{{\tilde{1}}} Y^{Q,RR}_{{\tilde{1}}} \right)
\end{autobreak}
\end{align}
\vspace{-0.5cm}
\begin{align}
\begin{autobreak}
\beta^{(2)}(g_2) =

+ \frac{9}{10} g_1^{2} g_2^{3}

+ \frac{35}{6} g_2^{5}

+ 12 g_2^{3} g_3^{2}

-  \frac{3}{2} g_2^{3} \tr\left(Y_u^{\dagger} Y_u \right)

-  \frac{3}{2} g_2^{3} \tr\left(Y_d^{\dagger} Y_d \right)

-  \frac{1}{2} g_2^{3} \tr\left(Y_e^{\dagger} Y_e \right)
\end{autobreak}
\end{align}

\vspace{-0.5cm}
\begin{align}
\begin{autobreak}
\beta^{(2)}(g_3) =

+ \frac{97}{30} g_1^{2} g_3^{3}

+ \frac{9}{2} g_2^{2} g_3^{3}

-  \frac{67}{3} g_3^{5}

- 2 g_3^{3} \tr\left(Y_u^{\dagger} Y_u \right)

- 2 g_3^{3} \tr\left(Y_d^{\dagger} Y_d \right)

-  \frac{1}{2} g_3^{3} \tr\left({Y^{RR}_{{\tilde{1}}}}^{\dagger} Y^{RR}_{{\tilde{1}}} \right)

+ 4 g_3^{3} \tr\left(Y^{Q,RR\,*}_{{\tilde{1}}} Y^{Q,RR}_{{\tilde{1}}} \right)
\end{autobreak}
\end{align}

\item {SM $ + \, \Phi_2$}
\begin{align}
\begin{autobreak}
\beta^{(2)}(g_1) =

+ \frac{1499}{75} g_1^{5}

+ \frac{87}{5} g_1^{3} g_2^{2}

+ \frac{524}{15} g_1^{3} g_3^{2}

-  \frac{17}{10} g_1^{3} \tr\left(Y_u^{\dagger} Y_u \right)

-  \frac{1}{2} g_1^{3} \tr\left(Y_d^{\dagger} Y_d \right)

-  \frac{3}{2} g_1^{3} \tr\left(Y_e^{\dagger} Y_e \right)

-  \frac{5}{2} g_1^{3} \tr\left({Y^{RL}_{2}}^{\dagger} Y^{RL}_{2} \right)

-  \frac{37}{10} g_1^{3} \tr\left({Y^{LR}_{2}}^{\dagger} Y^{LR}_{2} \right)
\end{autobreak}
\end{align}
\vspace{-0.5cm}
\begin{align}
\begin{autobreak}
\beta^{(2)}(g_2) =

+ \frac{29}{5} g_1^{2} g_2^{3}

+ \frac{37}{3} g_2^{5}

+ 20 g_2^{3} g_3^{2}

-  \frac{3}{2} g_2^{3} \tr\left(Y_u^{\dagger} Y_u \right)

-  \frac{3}{2} g_2^{3} \tr\left(Y_d^{\dagger} Y_d \right)

-  \frac{1}{2} g_2^{3} \tr\left(Y_e^{\dagger} Y_e \right)

-  \frac{3}{2} g_2^{3} \tr\left({Y^{RL}_{2}}^{\dagger} Y^{RL}_{2} \right)

-  \frac{3}{2} g_2^{3} \tr\left({Y^{LR}_{2}}^{\dagger} Y^{LR}_{2} \right)
\end{autobreak}
\end{align}
\vspace{-0.5cm}
\begin{align}
\begin{autobreak}
\beta^{(2)}(g_3) =

+ \frac{131}{30} g_1^{2} g_3^{3}

+ \frac{15}{2} g_2^{2} g_3^{3}

-  \frac{56}{3} g_3^{5}

- 2 g_3^{3} \tr\left(Y_u^{\dagger} Y_u \right)

- 2 g_3^{3} \tr\left(Y_d^{\dagger} Y_d \right)

-  g_3^{3} \tr\left({Y^{RL}_{2}}^{\dagger} Y^{RL}_{2} \right)

-  g_3^{3} \tr\left({Y^{LR}_{2}}^{\dagger} Y^{LR}_{2} \right)
\end{autobreak}
\end{align}
\item { SM $ + \, \Phi_{\tilde{2}}$}
\begin{align}
\begin{autobreak}
\beta^{(2)}(g_1) =

+ \frac{299}{75} g_1^{5}

+ 3 g_1^{3} g_2^{2}

+ \frac{28}{3} g_1^{3} g_3^{2}

-  \frac{17}{10} g_1^{3} \tr\left(Y_u^{\dagger} Y_u \right)

-  \frac{1}{2} g_1^{3} \tr\left(Y_d^{\dagger} Y_d \right)

-  \frac{3}{2} g_1^{3} \tr\left(Y_e^{\dagger} Y_e \right)

-  \frac{13}{10} g_1^{3} \tr\left({Y^{RL}_{{\tilde{2}}}}^{\dagger} Y^{RL}_{{\tilde{2}}} \right)
\end{autobreak}
\end{align}
\vspace{-0.7cm}
\begin{align}
\begin{autobreak}
\beta^{(2)}(g_2) =

+ g_1^{2} g_2^{3}

+ \frac{37}{3} g_2^{5}

+ 20 g_2^{3} g_3^{2}

-  \frac{3}{2} g_2^{3} \tr\left(Y_u^{\dagger} Y_u \right)

-  \frac{3}{2} g_2^{3} \tr\left(Y_d^{\dagger} Y_d \right)

-  \frac{1}{2} g_2^{3} \tr\left(Y_e^{\dagger} Y_e \right)

-  \frac{3}{2} g_2^{3} \tr\left({Y^{RL}_{{\tilde{2}}}}^{\dagger} Y^{RL}_{{\tilde{2}}} \right)
\end{autobreak}
\end{align}

\begin{align}
\begin{autobreak}
\beta^{(2)}(g_3) =

+ \frac{7}{6} g_1^{2} g_3^{3}

+ \frac{15}{2} g_2^{2} g_3^{3}

-  \frac{56}{3} g_3^{5}

- 2 g_3^{3} \tr\left(Y_u^{\dagger} Y_u \right)

- 2 g_3^{3} \tr\left(Y_d^{\dagger} Y_d \right)

-  g_3^{3} \tr\left({Y^{RL}_{{\tilde{2}}}}^{\dagger} Y^{RL}_{{\tilde{2}}} \right)
\end{autobreak}
\end{align}

\item { SM $ + \, \Phi_3$}
\begin{align}
\begin{autobreak}
\beta^{(2)}(g_1) =

+ \frac{207}{50} g_1^{5}

+ \frac{15}{2} g_1^{3} g_2^{2}

+ 12 g_1^{3} g_3^{2}

-  \frac{17}{10} g_1^{3} \tr\left(Y_u^{\dagger} Y_u \right)

-  \frac{1}{2} g_1^{3} \tr\left(Y_d^{\dagger} Y_d \right)

-  \frac{3}{2} g_1^{3} \tr\left(Y_e^{\dagger} Y_e \right)

- 3 g_1^{3} \tr\left({Y^{LL}_3}^{\dagger} Y^{LL}_3 \right)

+ \frac{12}{5} g_1^{3} \tr\left(Y^{Q,LL\,*}_3 Y^{Q,LL}_3 \right)
\end{autobreak}
\end{align}
\vspace{-0.6cm}
\begin{align}
\begin{autobreak}
\beta^{(2)}(g_2) =

+ \frac{5}{2} g_1^{2} g_2^{3}

+ \frac{371}{6} g_2^{5}

+ 44 g_2^{3} g_3^{2}

-  \frac{3}{2} g_2^{3} \tr\left(Y_u^{\dagger} Y_u \right)

-  \frac{3}{2} g_2^{3} \tr\left(Y_d^{\dagger} Y_d \right)

-  \frac{1}{2} g_2^{3} \tr\left(Y_e^{\dagger} Y_e \right)

- 9 g_2^{3} \tr\left({Y^{LL}_3}^{\dagger} Y^{LL}_3 \right)

+ 36 g_2^{3} \tr\left(Y^{Q,LL\,*}_3 Y^{Q,LL}_3 \right)
\end{autobreak}
\end{align}
\vspace{-0.6cm}
\begin{align}
\begin{autobreak}
\beta^{(2)}(g_3) =

+ \frac{3}{2} g_1^{2} g_3^{3}

+ \frac{33}{2} g_2^{2} g_3^{3}

- 15 g_3^{5}

- 2 g_3^{3} \tr\left(Y_u^{\dagger} Y_u \right)

- 2 g_3^{3} \tr\left(Y_d^{\dagger} Y_d \right)

- 3 g_3^{3} \tr\left({Y^{LL}_3}^{\dagger} Y^{LL}_3 \right)

+ 24 g_3^{3} \tr\left(Y^{Q,LL\,*}_3 Y^{Q,LL}_3 \right)
\end{autobreak}
\end{align}

\item { SM $ + \, \Phi_{2}+ \, \Phi_{3}$}
\begin{align}
\begin{autobreak}
\beta^{(2)}(g_1) =

+ \frac{1511}{75} g_1^{5}

+ \frac{111}{5} g_1^{3} g_2^{2}

+ \frac{572}{15} g_1^{3} g_3^{2}

-  \frac{17}{10} g_1^{3} \tr\left(Y_u^{\dagger} Y_u \right)

-  \frac{1}{2} g_1^{3} \tr\left(Y_d^{\dagger} Y_d \right)

-  \frac{3}{2} g_1^{3} \tr\left(Y_e^{\dagger} Y_e \right)

-  \frac{5}{2} g_1^{3} \tr\left({Y^{RL}_{2}}^{\dagger} Y^{RL}_{2} \right)

-  \frac{37}{10} g_1^{3} \tr\left({Y^{LR}_{2}}^{\dagger} Y^{LR}_{2} \right)

- 3 g_1^{3} \tr\left({Y^{LL}_3}^{\dagger} Y^{LL}_3 \right)

+ \frac{12}{5} g_1^{3} \tr\left(Y^{Q,LL\,*}_3 Y^{Q,LL}_3 \right)
\end{autobreak}
\end{align}

\begin{align}
\begin{autobreak}
\beta^{(2)}(g_2) =

+ \frac{37}{5} g_1^{2} g_2^{3}

+ \frac{205}{3} g_2^{5}

+ 52 g_2^{3} g_3^{2}

-  \frac{3}{2} g_2^{3} \tr\left(Y_u^{\dagger} Y_u \right)

-  \frac{3}{2} g_2^{3} \tr\left(Y_d^{\dagger} Y_d \right)

-  \frac{1}{2} g_2^{3} \tr\left(Y_e^{\dagger} Y_e \right)

-  \frac{3}{2} g_2^{3} \tr\left({Y^{RL}_{2}}^{\dagger} Y^{RL}_{2} \right)

-  \frac{3}{2} g_2^{3} \tr\left({Y^{LR}_{2}}^{\dagger} Y^{LR}_{2} \right)

- 9 g_2^{3} \tr\left({Y^{LL}_3}^{\dagger} Y^{LL}_3 \right)

+ 36 g_2^{3} \tr\left(Y^{Q,LL\,*}_3 Y^{Q,LL}_3 \right)
\end{autobreak}
\end{align}

\begin{align}
\begin{autobreak}
\beta^{(2)}(g_3) =

+ \frac{143}{30} g_1^{2} g_3^{3}

+ \frac{39}{2} g_2^{2} g_3^{3}

-  \frac{23}{3} g_3^{5}

- 2 g_3^{3} \tr\left(Y_u^{\dagger} Y_u \right)

- 2 g_3^{3} \tr\left(Y_d^{\dagger} Y_d \right)

-  g_3^{3} \tr\left({Y^{RL}_{2}}^{\dagger} Y^{RL}_{2} \right)

-  g_3^{3} \tr\left({Y^{LR}_{2}}^{\dagger} Y^{LR}_{2} \right)

- 3 g_3^{3} \tr\left({Y^{LL}_3}^{\dagger} Y^{LL}_3 \right)

+ 24 g_3^{3} \tr\left(Y^{Q,LL\,*}_3 Y^{Q,LL}_3 \right)
\end{autobreak}
\end{align}
\end{itemize}

\hypertarget{AppRGEYuk}{\section{One-Loop $\beta$-functions of SM Yukawa couplings \label{AppYukawa}}}
This Appendix presents the one-loop $\beta$-functions of the SM Yukawa couplings. We use  $n_1$, $n_{\tilde{1}}$, $n_2$, $n_{\tilde{2}}$ and $n_3$ to denote the number of generations of $\Phi_{1}$, $\Phi_{\tilde{1}}$, $\Phi_{2}$, $\Phi_{\tilde{2}}$ and $\Phi_{3}$, respectively.

\begin{align}
& \beta^{(1)}(Y_u) =
 \frac{3}{2} Y_u Y_u^{\dagger} Y_u
-  \frac{3}{2} Y_d Y_d^{\dagger} Y_u
+ 3 \tr\left(Y_u^{\dagger} Y_u \right) Y_u
+ 3 \tr\left(Y_d^{\dagger} Y_d \right) Y_u
+ \tr\left(Y_e^{\dagger} Y_e \right) Y_u \nonumber \\ 
&
-  \frac{17}{20} g_1^{2} Y_u
-  \frac{9}{4} g_2^{2} Y_u
- 8 g_3^{2} Y_u + n_1 \bigg(  \frac{1}{2} Y_u Y^{RR\,*}_1 {Y^{RR}_1}^{\trans}
+ Y_u Y^{Q,RR\,*}_1 {Y^{Q,RR}_1}^{\trans}
\nonumber \\ 
&
+ 2 Y^{LL\,*}_1 Y_e^{*} {Y^{RR}_1}^{\trans}
+ \frac{1}{2} Y^{LL\,*}_1 {Y^{LL}_1}^{\trans} Y_u
+ 8 Y^{Q,LL\,*}_1 Y_d^{*} {Y^{Q,RR}_1}^{\trans}
+ 4 Y^{Q,LL\,*}_1 Y^{Q,LL}_1 Y_u \bigg)
\nonumber \\ 
&
- n_{\tilde{1}} \bigg( 
4 Y_u Y^{Q,RR\,*}_{\tilde{1}} Y^{Q,RR}_{\tilde{1}}
\bigg)
+ n_2 \bigg(
Y_u Y^{RL}_{2} {Y^{RL}_{2}}^{\dagger}
+ 2 Y^{LR}_{2} Y_e^{\dagger} {Y^{RL}_{2}}^{\dagger}
+ \frac{1}{2} Y^{LR}_{2} {Y^{LR}_{2}}^{\dagger} Y_u
\bigg)
\nonumber \\ 
&
+ n_3 \bigg(
\frac{3}{2} Y^{LL\,*}_3 {Y^{LL}_3}^{\trans} Y_u
- 12 Y^{Q,LL\,*}_3 Y^{Q,LL}_3 Y_u
\bigg)
\end{align}
\begin{align}
& \beta^{(1)}(Y_d)=
\frac{3}{2} Y_d Y_d^{\dagger} Y_d
-\frac{3}{2} Y_u Y_u^{\dagger} Y_d
+ 3 \tr\left(Y_u^{\dagger} Y_u \right) Y_d
+ 3 \tr\left(Y_d^{\dagger} Y_d \right) Y_d
+ \tr\left(Y_e^{\dagger} Y_e \right) Y_d
-  \frac{1}{4} g_1^{2} Y_d
\nonumber \\ 
& -  \frac{9}{4} g_2^{2} Y_d
- 8 g_3^{2} Y_d
+ n_1 \bigg( 
Y_d {Y^{Q,RR}_1}^{\dagger} Y^{Q,RR}_1
+ \frac{1}{2} Y^{LL\,*}_1 {Y^{LL}_1}^{\trans} Y_d
+ 8 Y^{Q,LL\,*}_1 Y_u^{*} Y^{Q,RR}_1
\nonumber \\ 
& + 4 Y^{Q,LL\,*}_1 Y^{Q,LL}_1 Y_d
\bigg)
+ n_{\tilde{1}} \bigg( 
\frac{1}{2} Y_d Y^{RR\,*}_{\tilde{1}} {Y^{RR}_{\tilde{1}}}^{\trans} \bigg)
+ n_2 \bigg(
\frac{1}{2} Y^{LR}_{2} {Y^{LR}_{2}}^{\dagger} Y_d
\bigg)
+ n_{\tilde{2}} \bigg(
Y_d Y^{RL}_{\tilde{2}} {Y^{RL}_{\tilde{2}}}^{\dagger}
\bigg)
\nonumber \\ 
& + n_3 \bigg(
\frac{3}{2} Y^{LL\,*}_3 {Y^{LL}_3}^{\trans} Y_d
- 12 Y^{Q,LL\,*}_3 Y^{Q,LL}_3 Y_d
\bigg)
\end{align}
\begin{align}
\beta&^{(1)}(Y_e)=
\frac{3}{2} Y_e Y_e^{\dagger} Y_e
+ 3 \tr\left(Y_u^{\dagger} Y_u \right) Y_e
+ 3 \tr\left(Y_d^{\dagger} Y_d \right) Y_e
+ \tr\left(Y_e^{\dagger} Y_e \right) Y_e
-  \frac{9}{4} g_1^{2} Y_e
-  \frac{9}{4} g_2^{2} Y_e
\nonumber \\ 
&
+ n_1 \bigg( 
+ \frac{3}{2} Y_e {Y^{RR}_1}^{\dagger} Y^{RR}_1

+ 6 {Y^{LL}_1}^{\dagger} Y_u^{*} Y^{RR}_1

+ \frac{3}{2} {Y^{LL}_1}^{\dagger} Y^{LL}_1 Y_e
\bigg)+ n_{\tilde{1}} \bigg( 
\frac{3}{2} Y_e {Y^{RR}_{\tilde{1}}}^{\dagger} Y^{RR}_{\tilde{1}}
\bigg)
\nonumber \\ 
&
+ n_2 \bigg(
3 Y_e {Y^{LR}_{2}}^{\dagger} Y^{LR}_{2}
+ 6 {Y^{RL}_{2}}^{\dagger} Y_u^{\dagger} Y^{LR}_{2}
+ \frac{3}{2} {Y^{RL}_{2}}^{\dagger} Y^{RL}_{2} Y_e
\bigg)+ n_{\tilde{2}} \bigg(
\frac{3}{2} {Y^{RL}_{\tilde{2}}}^{\dagger} Y^{RL}_{\tilde{2}} Y_e
\bigg) \nonumber \\ 
&
+ n_3 \bigg(
\frac{9}{2} {Y^{LL}_3}^{\dagger} Y^{LL}_3 Y_e
\bigg)
\end{align}

\hypertarget{AppThreshold}{\section{One-Loop threshold corrections}}
In this Appendix, we collect the one-loop threshold corrections of the SM gauge, Yukawa, and quartic couplings on matching the SM with SM$+ \Phi_1 + \cdots + \Phi_3$. To simplify the expressions, we assumed all the LQs to have the same mass $m$ (for unequal mass, see Ref.~\cite{GitHubSLQ}) and set the dimensionful trilinear couplings to zero. The superscript $0$ denotes the original parameters in the SM lagrangian and we use the shorthand notation ${L}_{x m}= 1+ x \log \left( \frac{\mu^2}{m^2} \right).$

\begin{align}
g^{(0)}_1 = g_1-\frac{3 g_1^3 \log \left(\frac{\mu }{m^2}\right)}{32 \pi ^2},
\hspace{0.5cm} g^{(0)}_2 = g_2-\frac{3 g_2^3 \log \left(\frac{\mu }{m^2}\right)}{32 \pi ^2}, \hspace{0.5cm}
g^{(0)}_3 = g_3-\frac{3 g_3^3 \log \left(\frac{\mu }{m^2}\right)}{64 \pi ^2}
\end{align}
\begin{align}
    \lambda^{(0)} &= \lambda -\frac{3 Y_{\tilde{1} 3} Y_{\tilde{1}3}^* \log \left(\frac{\mu }{m^2}\right)}{8 \pi ^2}-\frac{3 Y_{2 \tilde{2}} Y_{2\tilde{2}}^* \log \left(\frac{\mu }{m^2}\right)}{16 \pi ^2}-\frac{3 Y_{\tilde{1}}^2 \log \left(\frac{\mu }{m^2}\right)}{32 \pi ^2}-\frac{3
   Y_{\tilde{2}}^2 \log \left(\frac{\mu }{m^2}\right)}{16 \pi ^2}
   \nonumber \\
   &
   -\frac{3 Y_{\tilde{2} \tilde{2}}^2 \log \left(\frac{\mu }{m^2}\right)}{32 \pi ^2}-\frac{3
   Y_{\tilde{2}} Y_{\tilde{2} \tilde{2}} \log \left(\frac{\mu }{m^2}\right)}{16 \pi ^2}-\frac{3 Y_{13} Y_{13}^* \log \left(\frac{\mu
   }{m^2}\right)}{16 \pi ^2} -\frac{3 Y_1^2 \log \left(\frac{\mu }{m^2}\right)}{32 \pi ^2}
   \nonumber \\
   &-\frac{3 Y_2^2 \log \left(\frac{\mu
   }{m^2}\right)}{16 \pi ^2}-\frac{9 Y_3^2 \log \left(\frac{\mu }{m^2}\right)}{32 \pi ^2}-\frac{3 Y_{22}^2 \log \left(\frac{\mu }{m^2}\right)}{32
   \pi ^2}-\frac{3 Y_{33}^2 \log \left(\frac{\mu }{m^2}\right)}{16 \pi ^2}-\frac{3 Y_2 Y_{22} \log \left(\frac{\mu }{m^2}\right)}{16 \pi ^2}
\end{align}

\begin{align}
    Y^{(0)}_u &=Y_u+ \frac{Y_u Y_{\tilde{1}}^{{Q,RR} \, \dagger } Y_{\tilde{1}}^{{Q,RR}^T}}{16 \pi^2} {L}_{2m}
   -\frac{Y_1^{{Q,LL}^*} Y_d^* Y_1^{{Q,RR}^T}}{4 \pi ^2} {L}_{m}
   -\frac{Y_1^{{LL}^*} Y_e^* Y_1^{{RR}^T}}{16 \pi ^2} {L}_{m}
   \nonumber \\
   &
   -\frac{Y_1^{{LL}^*} Y_1^{{LL}^T} Y_u}{128 \pi ^2} {L}_{2m}
   -\frac{3 Y_3^{{LL}^*} Y_3^{{LL}^T} Y_u}{128 \pi ^2} {L}_{2m}
   -\frac{Y_1^{{Q,LL}^*} Y_1^{{Q,LL}^T} Y_u}{16 \pi ^2} {L}_{2m}
   \nonumber \\
    &
    +\frac{3 Y_3^{{Q,LL}^*} Y_3^{{Q,LL}} Y_u}{16 \pi ^2} {L}_{2m}
   -\frac{Y_u Y_1^{{Q,RR}^*} Y_1^{{Q,RR}^T}}{64 \pi ^2} {L}_{2m}
   -\frac{Y_u Y_1^{{RR}^*} Y_1^{{RR}^T}}{128 \pi ^2} {L}_{2m}
   \nonumber \\
    &
    +\frac{Y_2^{LR} Y_e^{\dagger} Y_2^{{RL}^\dagger}}{16 \pi ^2} {L}_{m}
   -\frac{Y_2^{{LR}} Y_2^{{LR}^\dagger} Y_u}{128 \pi ^2} {L}_{2m}
   -\frac{Y_u Y_2^{RL} Y_2^{{RL}^\dagger}}{64 \pi ^2} {L}_{2m}
\end{align}

\begin{align}
    Y^{(0)}_d&=Y_d
    -\frac{Y_d Y_{\tilde{1}}^{{RR}^*} Y_{\tilde{1}}^{{RR}^T}}{128 \pi ^2} {L}_{2m}
   -\frac{Y_d Y_{\tilde{2}}^{{RL}} Y_{\tilde{2}}^{{RL}^{\dagger}}}{64 \pi ^2} {L}_{2m}
   -\frac{Y_{1}^{{LL}^*} Y_{1}^{{LL}^{T}} Y_d}{128 \pi ^2} {L}_{2m}
   \nonumber \\ 
   &-\frac{3 Y_{3}^{{LL}^*} Y_{3}^{{LL}^{T}} Y_d}{128 \pi ^2} {L}_{2m}-\frac{Y_{1}^{{Q,LL}^*} Y_{1}^{{Q,LL}^{T}} Y_d}{16 \pi ^2} {L}_{2m}
   +\frac{3 Y_{3}^{{Q,LL}^*} Y_{3}^{{Q,LL}} Y_d}{16 \pi ^2} {L}_{2m}
   \nonumber \\ 
   &-\frac{Y_{1}^{{Q,LL}^*} Y_u^* Y_{1}^{{Q,RR}}}{4 \pi ^2} {L}_{m}
   -\frac{Y_{2}^{{LR}} Y_{2}^{{LR}^\dagger} Y_d}{128 \pi ^2} {L}_{2m}
   -\frac{Y_d Y_{1}^{{Q,RR}^\dagger} Y_{1}^{{Q,RR}}}{64 \pi ^2} {L}_{2m}
\end{align}

\begin{align}
    Y^{(0)}_e &= Y_e -\frac{3 Y_{\tilde{2}}^{{RL}^\dagger} Y_{\tilde{2}}^{{RL}} Y_{e}}{128 \pi^2} {L}_{2m}
  -\frac{3 Y_{e} Y_{\tilde{1}}^{{RR}^\dagger} Y_{\tilde{1}}^{{RR}}}{128 \pi^2} {L}_{2m}  -\frac{3 Y_{1}^{{LL}^\dagger} Y_u^* Y_{1}^{{RR}}}{16 \pi^2} {L}_{m}
    \nonumber \\
    & -\frac{3 Y_{1}^{{LL}^\dagger} Y_{1}^{{LL}} Y_{e}}{128 \pi^2} {L}_{2m}
    -\frac{9 Y_{3}^{{LL}^\dagger} Y_{3}^{{LL}} Y_{e}}{128 \pi^2} {L}_{2m}
   -\frac{3 Y_{e} Y_{2}^{{LR}^\dagger} Y_{2}^{{LR}}}{64 \pi^2} {L}_{2m}
   \nonumber \\
   & -\frac{3 Y_{2}^{{RL}^\dagger} Y_{2}^{{RL}} Y_{e}}{128 \pi^2} {L}_{2m}
   -\frac{3 Y_{e} Y_{1}^{{RR}^\dagger} Y_{1}^{{RR}}}{128 \pi^2} {L}_{2m}+\frac{3 Y_{2}^{{RL}^\dagger} Y_{u}^{\dagger} Y_{2}^{{LR}}}{16 \pi^2} {L}_{m}
\end{align}

\bibliographystyle{JHEP}
\bibliography{ref}
\end{document}